\documentclass[letterpaper,12pt]{article}

\usepackage{natbib}
\usepackage{graphicx}
\usepackage{color}
\usepackage{hyperref}
\usepackage{lineno}
\usepackage{amsmath}
\usepackage{braket}
\usepackage{bm}
\usepackage{amssymb}
\usepackage{multirow}
\usepackage{lineno}
\usepackage{setspace}
\usepackage[normalem]{ulem}

\usepackage{authblk}

\usepackage{mathrsfs}
\usepackage{wasysym}
\usepackage{lineno}
\usepackage{epstopdf}

\begin{document}
\title{Explaining the Variations in Isotopic Ratios in Meteoritic Amino Acids}
	
\author[1,2,3]{Michael A. Famiano\footnote{Corresponding Author: \href{mailto:michael.famiano@wmich.edu}{michael.famiano@wmich.edu}, 269-387-4931}}
\author[4]{Richard N. Boyd\footnote{\href{mailto:richard11boyde@comcast.net}{richard11boyde@comcast.net}}}
\author[2,3,5]{Toshitaka Kajino\footnote{\href{mailto:kajino@nao.ac.jp}{kajino@nao.ac.jp}}}
\author[3,6]{Satoshi Chiba\footnote{\href{mailto:chiba.satoshi@nr.titech.ac.jp}{chiba.satoshi@nr.titech.ac.jp}}}
\author[7]{
Yirong Mo\footnote{\href{mailto:yirong.mo@wmich.edu}{yirong.mo@wmich.edu}}
}
\author[8,9]{Takashi Onaka\footnote{\href{mailto:onaka@astron.s.u-tokyo.ac.jp}{onaka@astron.s.u-tokyo.ac.jp}}}
\author[2,10]{Toshio Suzuki
\footnote{\href{mailto:suzuki@phys.chs.nihon-u.ac.jp}{suzuki@phys.chs.nihon-u.ac.jp}}
}

\affil[1]{\small Dept. of Physics and Joint Institute for Nuclear Astrophysics, Western Michigan University, 
	1903 W. Michigan Avenue, Kalamazoo, MI 49008-5252, USA}
\affil[2]{National Astronomical 
	Observatory of Japan, 
	2-21-1 Mitaka, Tokyo 181-8588 Japan}
\affil[3]{School of Physics, Beihang Univ. (Beijing Univ. of Aeronautics and Astronautics), Beijing 100083, P.R. China}
\affil[4]{Department of Physics, Department of Astronomy,
The Ohio State University, Columbus, OH 43210 USA}
\affil[5]{Graduate School of Science,
Univ. of Tokyo, 7-3-1 Hongo, Bukyo-ku, Tokyo 113-0033 Japan}
\affil[6]{Laboratory for Advanced Nuclear Energy, 
	Institute of Innovative Research, Tokyo Institute of Technology,
2-12-1-N1-16 Ookayama, Meguro-ku, Tokyo, 152-8550, Japan}
\affil[7]{Department of Chemistry, Western Michigan University,
Kalamazoo, MI 49008-5252 USA}
\affil[8]{Dept. of Astronomy, 
	Graduate School of Science; Univ. of Tokyo, 7-3-1 
	Hongo, Bunkyo-ku, Tokyo 113-0033 Japan}
\affil[9]{Dept. of Physics, Meisei University,  2-2-1 Hodokubo, Hino, Tokyo 191-8506, Japan}
\affil[10]{Department of Physics, Nihon University, 3-25-40, Setagaya-ku, Tokyo 156-8550, Japan}

\maketitle 
\newpage
\begin{abstract}
 Measurements of the isotopic abundances in meteoritic amino acids have found  enhancements of $^2$H/H, $^{15}$N/$^{14}$N, and $^{13}$C/$^{12}$C 
 in the amino acids in the meteorites studied. We show that they are consistent with the processing of the constituents of the meteorites by electron anti-neutrinos that would be expected from a core-collapse supernova or neutron-star merger. Using theoretical electron anti-neutrino cross sections we are able to 
 predict these isotopic ratio variations depending on
 the time-integrated anti-neutrino flux at the site where the amino acids were processed.  
\end{abstract}


\doublespacing
\section{Introduction} 
One of Nature's curiosities is the isotopic abundances observed in the constituents of 
the amino acids found in meteorites. Specifically, $^{13}$C is enhanced slightly over its
cosmic abundance, $^{15}$N considerably more, and $^2$H is greatly enhanced. 
An additional feature is the fact that these amino acids tend to be slightly skewed toward left-handedness. Since this is as
observed nearly in totality in Earthly amino acids (except for the achiral glycine) this has been used to suggest that meteorites may have provided the seeds for the ultimate production of 
Earthly amino acids.

In earlier work we studied a model by which the left-handedness might have been produced. In this paper we extend that model to study how isotopic ratios variations may be created in meteorites.

Several explanations have been given as to the origin of cosmic handedness 
selection \citep[e.g.~][]{shosuke10, breslow11, sojo15, gleiser12, kobayashi19, islas04,
podlech01, boyd18, uwe08}.  In this paper we observe that one model, the Supernova
Neutrino Amino Acid Processing (SNAAP) Model, can explain both the handedness selection
and the differences in isotopic abundances in a self-consistent way. Since the SNAAP Model's
handedness selection capability has been the subject of many 
papers~\citep{snaap1,snaap2,snaap3,snaap4,snaap5,snaap6}, this work will focus on its
ability to explain the differences in isotopic abundances.

In the next section we give a brief description of the SNAAP model, including a short
discussion of some of its results for amino acid handedness. Section 3
describes our calculations of the isotopic ratios, first describing that for $^2$H,
then using those results to determine the level of variation for $^{13}$C and $^{15}$N.We also mention the possibility of Li and B abundance ratios in the same meteorites.  

At the end of the results section, we evaluate some of the uncertainties in our
computations.  Results of our computations are compared to data
for meteoroids assuming isotopic abundance distributions corresponding to both the current solar system isotopic abundance distributions and the solar system
isotopic abundance distribution 4.6 billion years ago (Gya), where isotopic abundances were taken from \citet{lodders09}. We show that an evaluation of meteoroids with abundance distributions matching those of the solar system 4.6 Gya gives
the most reasonable agreement with observed values of ratios in D/H, $^{15}$N/$^{14}$N, and $^{13}$C/$^{12}$C ratios. The reason for this 
may be because meteoric isotopic abundances \textit{prior to} irradiation likely more closely matched those of the early solar system.

Finally we give our conclusions.  Here, we provide future perspectives and plans
as well as well as an iteration of the importance of sample return missions.  In addition,
we note the importance and past work in low-temperature chemistry (particularly 
D enhancement) \citep{elsila12} in interpreting measured values.  The model presented
here may complement existing interpretations.
\section{The SNAAP Model}
In this model \citep{snaap1,snaap2,snaap3,snaap4,snaap5,boyd18,snaap6}, meteoroids might be processed in the intense magnetic field and electron anti-neutrino (hereafter denoted anti-neutrino) flux from one of several stellar objects. The anti-neutrinos are selective in their destruction of the amino acids as they are purely chiral (assuming massless anti-neutrinos), a result of the weak interaction nuclear physics that describes their interaction with the $^{14}$N nuclei. The relevant nuclear reaction is
\begin{equation}
\bar{\nu}_e + ^{14}N \rightarrow ^{14}C + e^+	
\end{equation}
where $\bar{\nu}_e$ is an electron anti-neutrino and $e^+$ is an anti-electron, a 
positron. If the $\bar{\nu}_e$ spin (1/2, in units of $\hbar$, Planck’s constant divided 
by 2$\pi$) is antiparallel to that of the $^{14}$N (spin 1), then a total spin on the 
left-hand side of the equation is either 1/2 or 3/2. The smaller value will equal the sum
of the spins of $^{14}$C (spin 0) and the positron (spin 1/2) on the right-hand side, 
thus conserving angular momentum. However, if the $\bar{\nu}_e$ spin and the $^{14}$N 
spin are aligned, conservation of angular momentum will require one additional unit of 
angular momentum to come from either the $\bar{\nu}_e$ or the $e^+$ wave function in 
order for the total angular momentum on the right-hand side to equal the 3/2 on the 
left-hand side. This is known from basic nuclear physics \citep{snaap2} to introduce 
roughly an order of magnitude smaller cross section for the latter case compared to the 
former. However, this also introduces a parity change. Since the transition from $^{14}$N
to the $^{14}$C ground state is between nuclear states of the same parity, two units of 
angular momentum must therefore come from the anti-neutrino and/or positron wave
functions. Thus the inhibition may be closer to two orders of magnitude. This 
spin-dependent reaction probability is the
origin of the effect predicted for the SNAAP model.

Detailed quantum molecular calculations \citep{snaap4,snaap2} have shown that the complex
interactions of the molecules with the intense magnetic field of a neutron star or a two neutron star merger and the effective electric field caused by the motion of the meteoroids through the magnetic field, a result of the Lorentz force, do produce an environment that is truly chiral \citep{barron08}. In this situation, the interaction of the $^{14}$N nuclei with the $\bar{\nu}_e$s is chirally selective, and will in many cases, destroy more of the right-handed amino acids than the left-handed ones \citep{snaap4}.

The external magnetic field aligns the $^{14}$N nuclei via their nuclear magnetic moments,
whereas the effective electric field aligns the molecular electric dipole moments, which
depend on the chirality. The external magnetic field, however, is modified at the nucleus
by the effects of the orbital electrons, known as shielding. This, in turn causes a
chirality sensitive perturbation on the magnetic orientation of the nuclei, which leads to
a chirality-dependent magnetization (a bulk property).

These components exist even without the coupling to the electric dipole moment 
\citep{buckingham06,snaap4}, but that coupling enhances the difference between the angles 
that the two chiral states make with the anti-neutrino spin, hence of the chirality 
selective destruction of the amino acids \citep{snaap2,snaap4}. From the magnitude of these
effects, one can determine the enantiomeric excesses, $ee$s, (ee = (N$_L$-N$_D$)/(N$_L$+N$_D$), where N$_L$ (N$_D$) 
is the number of left-handed (right-handed) amino acids in any ensemble) that might be 
expected for amino acids from the SNAAP model. $Ee$s on the order of
percents are predicted for a variety of parameters in this model.

At least two sites seem promising for producing the effects we have studied: a supernova from a star that is in a massive star-neutron star binary and a neutron star-neutron star (NS) merger. Both could produce the anti-neutrino flux and magnetic field needed to process nearby objects such as large meteoroids or planets (the anti-neutrinos would process the entire object regardless of the size). Finally large objects passing by or in highly elliptical orbits would be necessary to limit the exposure to the high temperature near the central object. These have been discussed in previous work \citep{boyd18b,boyd18}.

Since life began on Earth more than 4 billion years ago, the detritus from such an event would have to have gotten to earth prior to that time. A supernova could have both created the processed amino acids and a shock wave that precipitated the solar system. A supernova subsequent to formation of the Solar System could also have sent amino acid laden debris to Earth. Such an event apparently deposited $^{60}$Fe (half-life of 1.5 Myr) on Earth roughly one million years ago \citep{koll19}, suggesting such depositions might well have provided the appropriately processed meteorites.   
 
In principle, electron neutrinos could drive the $^{14}$N to $^{14}$O, but the threshold
energy is much higher for this reaction (greater than 5 MeV compared to much less than 1
MeV). Since the cross section for neutrino capture processes increases sharply with the
energy above threshold, and the anti-neutrino energies expected \citep{rosswog03} from coalescing
neutron stars are predicted to be much larger for anti-neutrinos than neutrinos (16 MeV
versus 10 MeV), this reaction does not have a large effect on the enantiomerism that
results from the combined flux from anti-neutrinos and neutrinos in that site.
\section{Materials and Methods}
\subsection{Estimating Isotopic Ratios in Amino Acids}
The variations in isotopic abundances in the nuclei of the amino acids were estimated by assuming initial mass fractions equal to those found in
solar system meteoric mass fractions taken from Lodders (2009).  Initial mass
fractions were used for the solar system average, the Orgueil meteorite, and
the solar system at formation (4.6 Gya).  Exposure to an anti-neutrino flux
from a nearby source was then assumed.  The neutrino energies and
temperatures used for three models are indicated in Table 
\ref{models}.
These values were taken from the
Rosswog and Liebend\"orfer \citep{rosswog03}
evaluation of NS mergers and are representative of
typical neutrino luminosity and temperatures in such an event.

Neutrino and anti-neutrino charged-current (CC) and neutral-current (NC) interactions, as well as electron captures and $\beta$ decays,
have been included in a simple network to compute the changes in isotopic abundance as a function of time for a neutrino
flux from a given source.  

The isotopic abundance enhancements or deficiencies are evaluated by comparing to  
standards (VPDB for $^{13}$C~\citep{hut87}, 
VSMOW for $D$~\citep{hagemann70,dewit80}, and air for
$^{15}N$~\citep{junk58}) used in meteoric analysis.
The standard $\delta$ notation used to evaluate isotopic abundance is:
\begin{equation}
\label{ratio_eq}
\delta^A \equiv \left[\frac{R(A)_{sample}}{R(A)_{standard}}-1\right]\times 1000 \permil
\end{equation}
where $R(A)$ is the abundance ratio of an isotope of mass A to the most abundant isotope of the same element. 

For the present calculations, the reaction network was run for up to 10 s.  The values of $\delta$
were computed at each time step in the evaluation.  
\subsection{Model Parameters}
\subsubsection{Meteoroid Parameters}
\label{parameters}
The initial elemental abundances in the model meteoroid correspond to the CI abundances of the Orgueil 
meteorite rock average as stated in \citet{lodders09} with isotopic 
distributions for individual elements taken to be that of the solar
system 4.6 billion years ago (Table 10 from the same reference).  
Additionally, two other 
abundance distributions were chosen to match the abundance distribution of the solar system at formation 
\citep{lodders09} and the currently suggested solar photospheric values 
\citep{lodders09}.  

Although amino acids within meteoric inclusions are
not necessarily homogeneously distributed, it is safe to assume
a homogeneously mixed meteoroid for the following 
reasons. Here, we use the term 
\textit{inclusion} to indicate any inhomogeneity within 
the parent body, including cavities and pockets of 
water.  This is a more general definition than
in some literature \citep{chizmadia02,brearley99,pizzarello10}.  
In the current work, we are assuming that any
inclusion is $\lesssim$ 1 mm in size, similar in
size to grains and chondrules studied in carbonaceous
chondrites \citep{chizmadia02,brearley99,pizzarello10}.  The anti-neutrinos will penetrate and process the
entire meteorite evenly.  Any neutrons that are produced will have an initial mean free path of
a few centimeters, which is much larger than the expected size of the inclusions.  For 
this reason, neutrons produced in the vicinity of the inclusions will penetrate the 
volume of the meteoroid.  Neutrons produced within and without the inclusions were thus 
assumed to be uniformly distributed in energy and space. 
\subsubsection{Nuclear Reaction Network}
\begin{figure}[h!]
	\includegraphics[width=\textwidth]{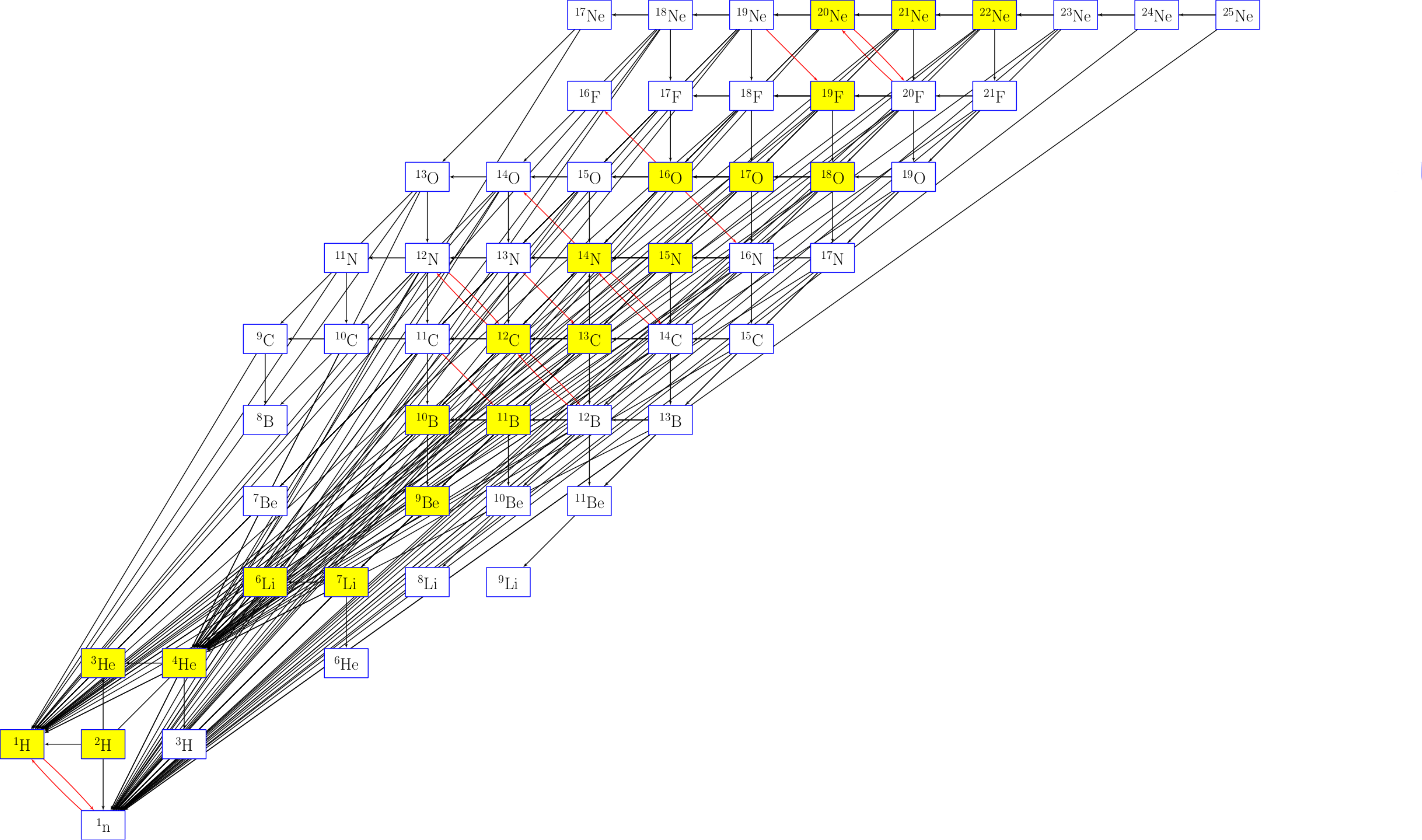}
	\caption{\label{w_net}The weak interaction network used to evaluate isotopic abundance distributions in amino acid formation.}
\end{figure}

Two reaction networks were used in this analysis.  Neutrino CC and NC reactions 
were evaluated using the network of Figure \ref{w_net} \citep{meyer12}.  
Each black line in the figure is a possible neutrino capture
reaction path.  For example, the reaction  
$^{12}$C+$\nu_\mu\rightarrow^{11}$C+n+$\nu_\mu^\prime$ would be represented by a black 
line.  The dominant $\beta^\pm$ decays are represented by red lines.  Stable 
nuclei are shaded in yellow. After 
the neutrino burst, post-processing by neutrons
produced via anti-neutrino captures on hydrogen and NC reactions was
incorporated into a network of species with Z$\le$26.  We base this choice on
the assumption that the environment was cold enough and sparse enough that only neutrino and neutron capture reactions occurred, and they did so at low
enough rates that the neutron captures were decoupled from the neutrino captures.  
Under these conditions, it is unlikely that any particular species will undergo multiple sequential
captures.

The anti-neutrino luminosities and temperatures specified in Table \ref{models}
were presumed constant throughout the duration of the burst.  Neutrino 
NC and CC cross sections were taken from the literature \citep{kolbe03,mmf98,strumia03,suzuki18,woosley90} where available.  For the
low-mass Li and B isotopes, cross-sections were evaluated using the
shell-model calculations of \citet{suzuki06}.
\subsubsection{Neutron Captures and Post-Processing}
Subsequent neutron captures are energy-dependent.  Neutron captures on all isotopes present in the meteoroid up to $^{56}$Fe were evaluated and were taken from
the ENDF compilation (ENDF/B-VIII) \citep{endf}. Here, the fraction
of neutrons $F_i$ captured on a species $i$ during thermalization is:
\begin{equation}
\label{neut_therm}
F_i = \frac{n_i\int_{E_\circ}^{\langle E \rangle} \sigma_i(E)\phi_{IBD}(E)dE}{\sum\limits_i n_i\int_{E_\circ}^{\langle E \rangle} \sigma_i(E)\phi_{IBD}(E)dE}
\end{equation}
where $\langle E\rangle$ is the average energy of the produced neutrons, and
$n_i$ is the number density of species $i$.  The integration is
over the entire energy spectrum of the neutrons as they thermalize to
a presumed lower energy $E_\circ$, which is taken to be 25 meV, low enough such that the cross-sections of all neutrons at this energy are
dominated by non-resonant capture.  The neutron energy spectrum, $\phi_{IBD}(E)$,
is taken to be that of proton inverse $\beta$ decay (IBD), which is determined
based on the energy distribution of neutrons produced in IBD and thermalized
in the intervening meteoric rock.

This assumes that the neutron scattering cross sections are dominant in the thermalization process.  Otherwise, a correction to the
integrals would require a shift in the neutron energy-dependent flux as neutrons are also captured as they thermalize.  However, since most of 
the captures occur at lower energies during the thermalization process, where the
relative rates scale only by the thermal neutron capture cross sections, the 
effect of the treatment given by Equation \ref{neut_therm} is to correct for
resonances at higher energy, where the capture cross section is much
less than the scattering cross section.
  
The energy loss rate $\frac{dE}{dx}$ is taken from the inelastic scattering
cross section of neutrons in material.  
From this, the energy loss rate can be calculated assuming
a constant scattering cross section of $\approx$5 b and an average energy
loss per collision of $f$=0.9 of the original energy \citep{krane}.  The
energy loss rate is then given by:
\begin{eqnarray}
E &= E_\circ f^{x/\lambda}
\\\nonumber
~&=E_\circ e^{\alpha x}
\\\nonumber
\rightarrow \frac{dE}{dx} &\propto e^{\alpha x}
\end{eqnarray}
where $\lambda$ is the mean free path of neutrons in the material.  In this
case, $\lambda$ is estimated to be 5 cm, although it depends on the composition
and density of the material.  As mentioned in \S\ref{parameters}, we can take this
as an average over the entire meteoroid.  While the energy loss rate
may be higher for aqueous inclusions, the inclusions are so small that the
neutrons will not thermalize significantly within them.  That is, $\lambda_n\gg r_i$,
where $\lambda_n$ is the neutron mean-free-path (in either water or rock) and 
$r_i$ is the inclusion radius.  With this condition satisfied, the meteoroid may be
treated as homogeneous.

In order to properly evaluate the neutron average energy used in Equation \ref{neut_therm}, the energy distribution of
neutrons from the NC and proton inverse beta decay must be calculated.
For the proton inverse beta decay, nearly all of the excess kinetic energy is
carried by the positron.  For the electron anti-neutrino energies 
listed in Table \ref{models}, the neutron energy is roughly
200 keV \citep{strumia03}.  The produced neutrons will then penetrate the surrounding medium, including the rock and any 
intervening organic material, losing energy as they proceed.

Nearly 100\% of all of the neutrons produced in the
reaction network come from inverse beta decay of the proton, given
that the charged-current cross section is several orders of magnitude larger
than the neutral current nuclear cross sections evaluated and the H abundance in the meteoric medium is significant.  Thus, we assumed average neutron energies corresponding to those emitted in charged-current reactions with hydrogen.
\section{Results}
\begin{figure}
	\centering
	\includegraphics[width=0.49\textwidth]{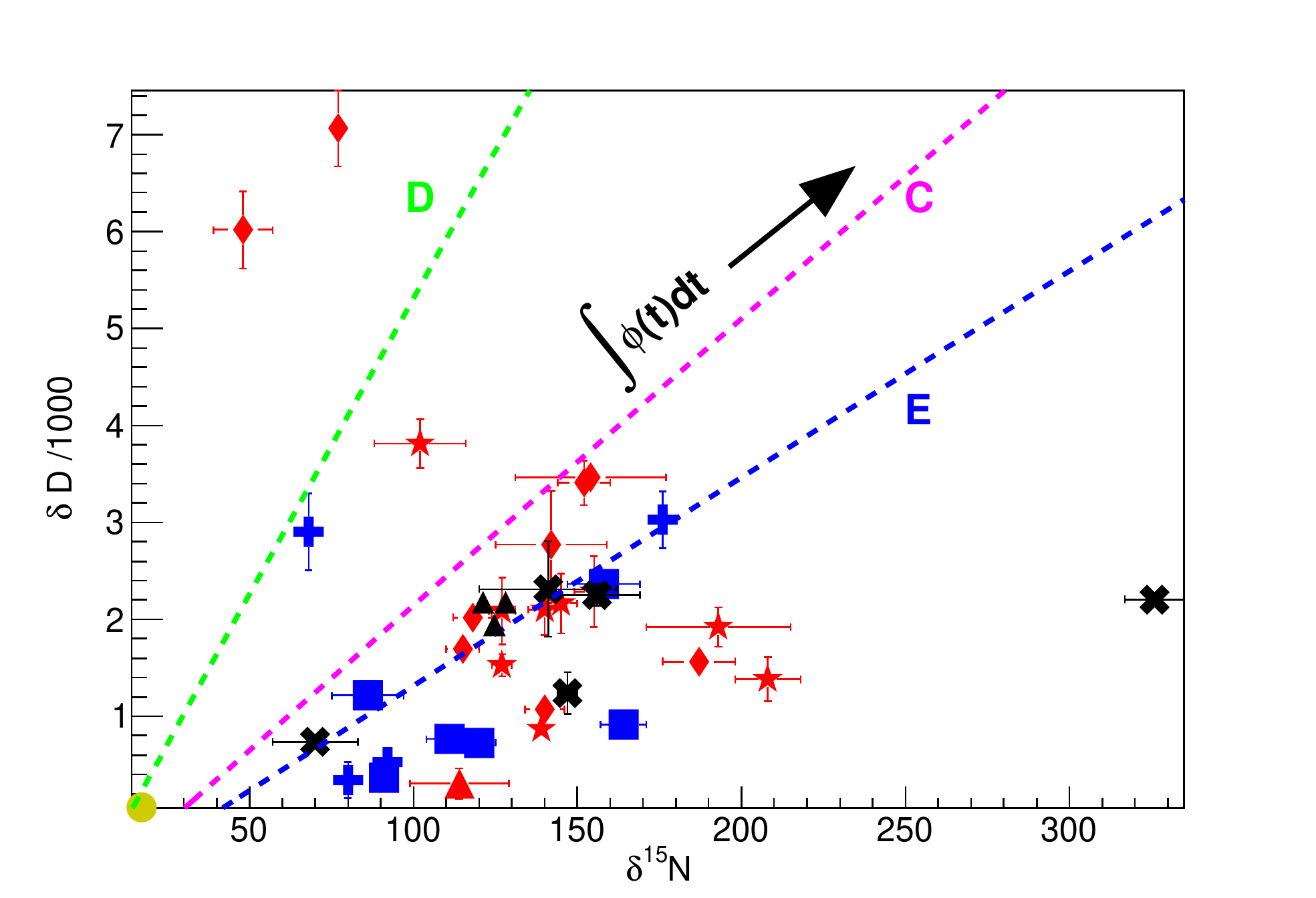}
		\includegraphics[width=0.49\textwidth]{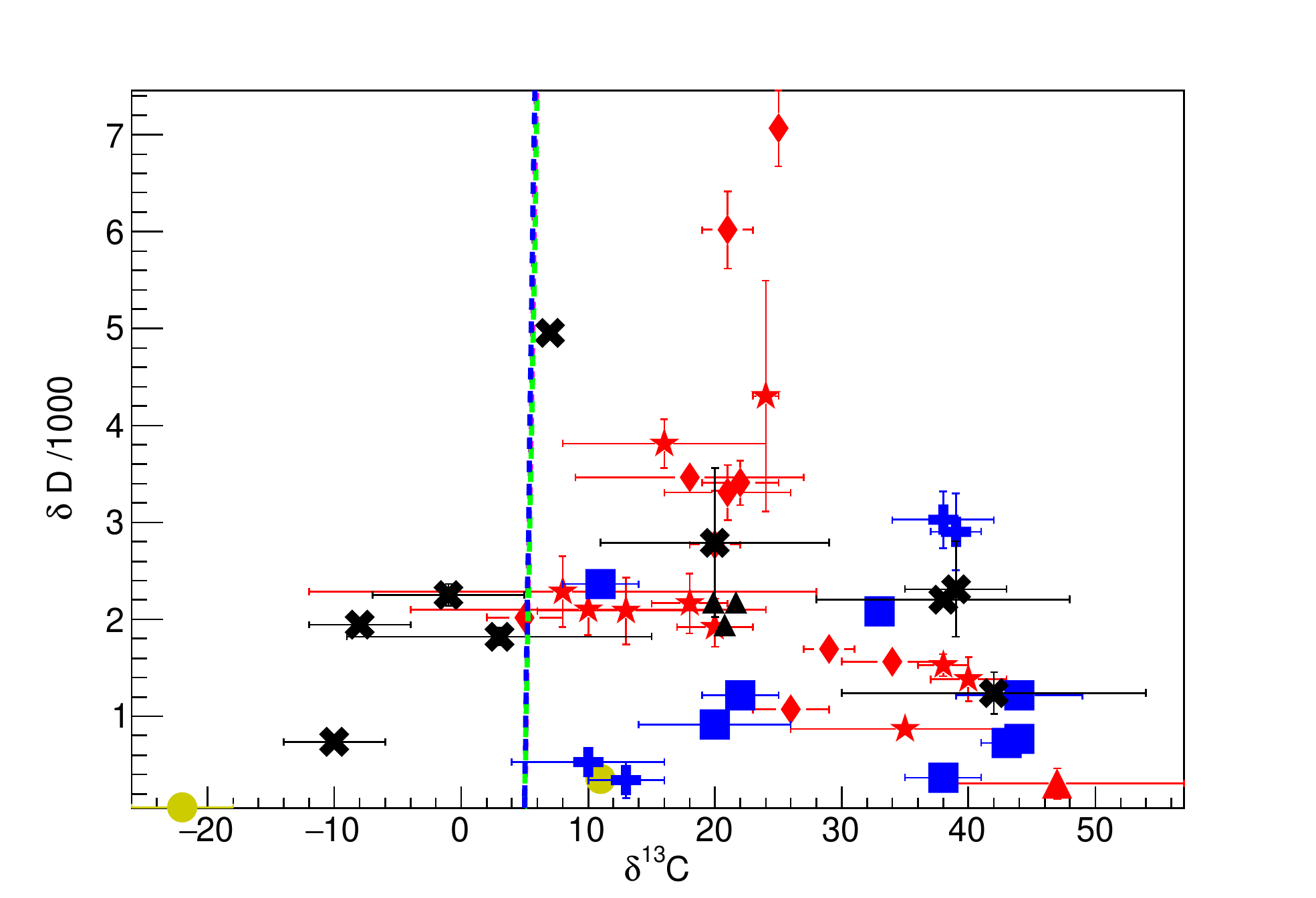}\\
			\includegraphics[width=0.49\textwidth]{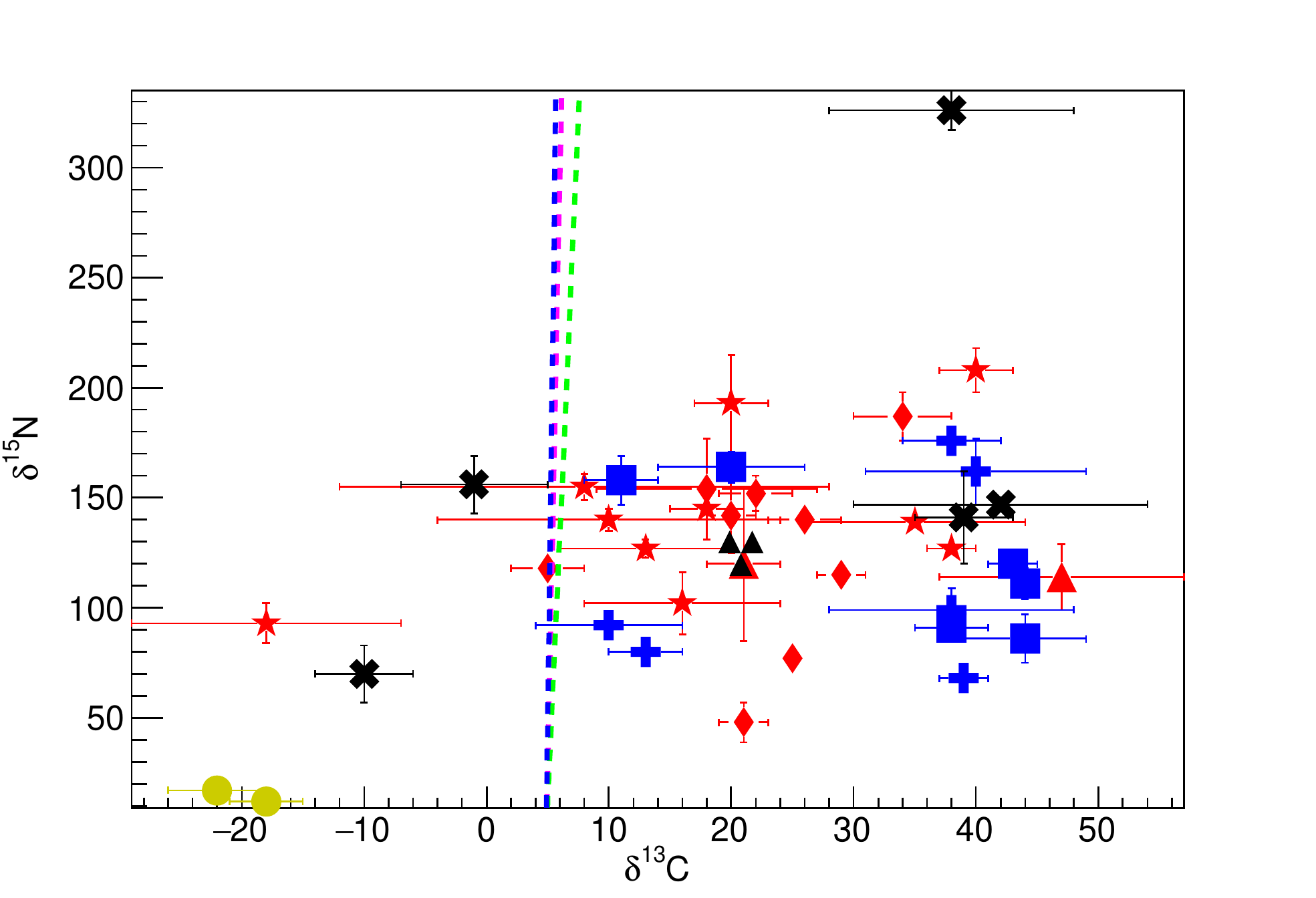}
				\includegraphics[width=0.49\textwidth]{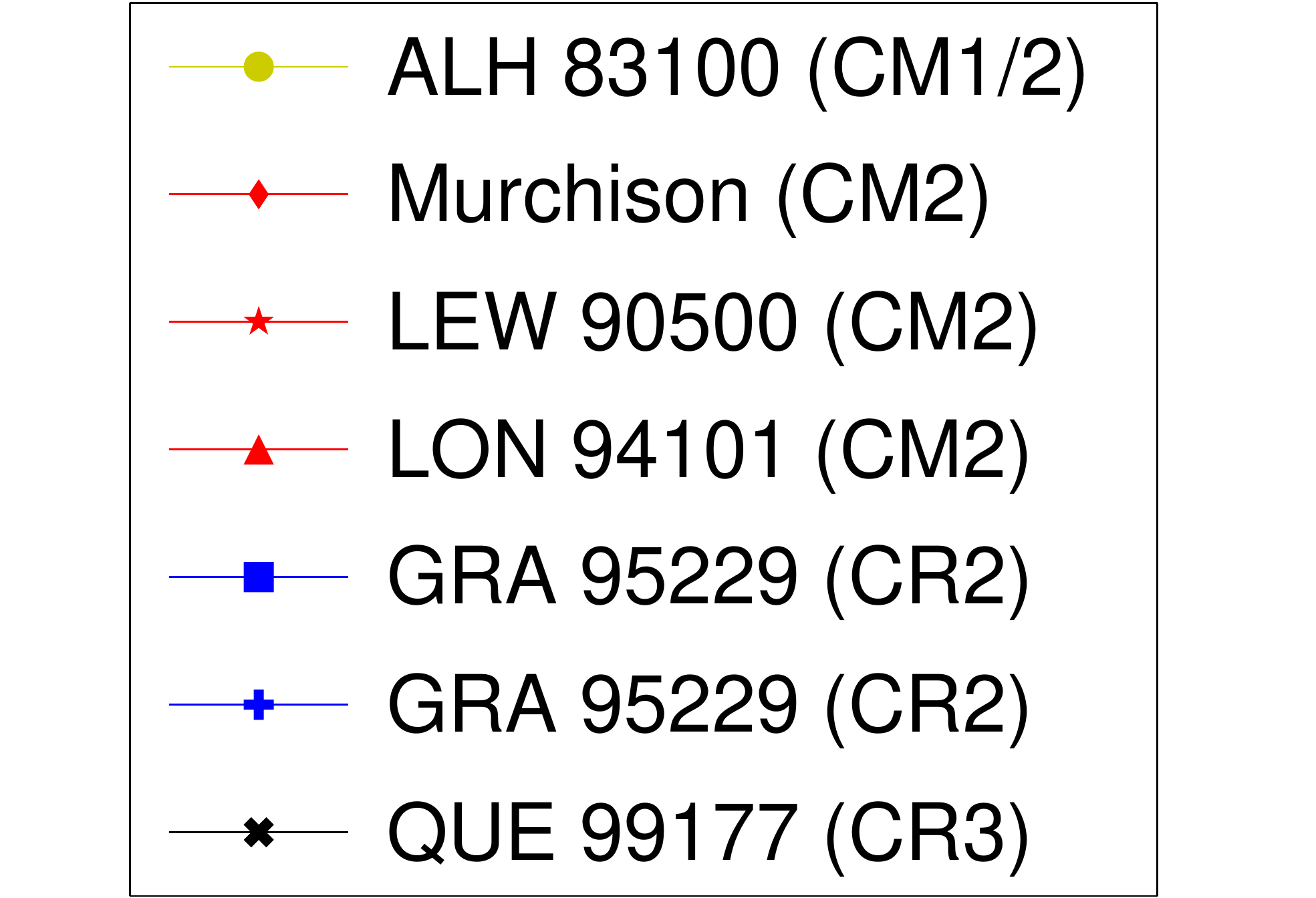}
	\caption{\label{anom_CI}
		The abundance ratios compared to the measurements of \citep{elsila12}.  Initial abundances
		are taken to be those of the Orgueil meteorite \citep{lodders09} with isotopic fractions 
		taken from the same reference.  The 
		symbols correspond to the results of Elsila, while the green, purple, and blue lines correspond
		to the results of the network calculation for models D, C, and E respectively.   
		The source meteorite for each measurement is indicated
		by the symbol type, and the meteorite type is indicated
		by the symbol color.  
		The 
		direction of increasing integrated neutrino fluence is also shown as indicated. The black trefoil in each figure represents the arithmetic mean of the data.			
	}
\end{figure}

Isotopic abundances have been determined for all species in the reaction network.  Comparisons have been 
made to the meteoric measurements of Elsila et al. \citep{elsila12}.  The results are shown
in Figures \ref{anom_CI} - \ref{anom_S}. 

In all figures, the red, green, and purple lines show the calculated results based on models C, D, and E of Table \ref{models}, where 
each curve is a parametric line in time 
(corresponding to increasing neutrino fluence).  That is, for each line in the figure, neutrino exposure time increases from the
lowest values of the isotopic ratios to the highest values.
The direction of increasing exposure time (fluence) is indicated 
in Figure \ref{anom_CI}.  
Because a constant neutrino flux is assumed, increasing time corresponds to 
increasing integrated flux (fluence):
\begin{equation}
\mathscr{F}(t)\equiv\int\phi(t) dt
\end{equation}
As the exposure time and fluence are variable, any point on the lines in the figures represents a valid possible set of isotopic ratios.

This is important to note for the following reason.  While the computation assumes constant flux, for the strong neutrino flux assumed, the neutrino interactions dominate over other reactions
in the short timescale noted here.  In evaluating the values of $\delta$, we have assumed that all
unstable nuclei have decayed back to stability.  This means that the points at which the trends indicated by the calculated lines in the figures show the final equilibrium value of $\delta$ at a particular fluence.

The measured values in each figure are indicated by the markers.  Individual marker types correspond to the meteorite from which the amino acids were extracted.  The color of the marker corresponds to the type of meteorite
from which the amino acids were extracted.  Red, blue, black, and dark yellow markers correspond to CM2, CR2, CR3, and CM1/2 meteorites respectively. 
\section{Discussion}
\begin{figure}
	\centering
	\includegraphics[width=0.49\textwidth]{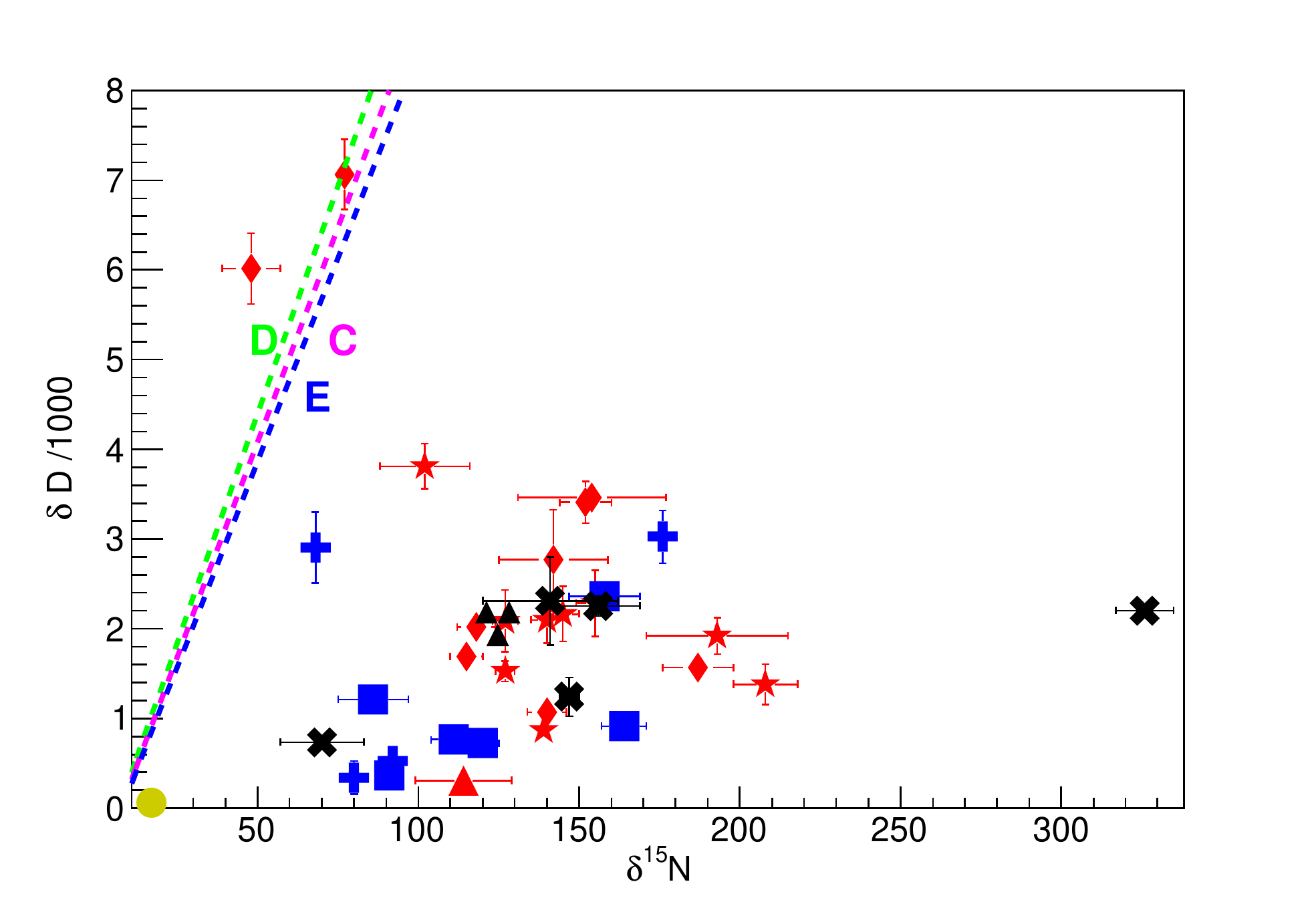}
	\includegraphics[width=0.49\textwidth]{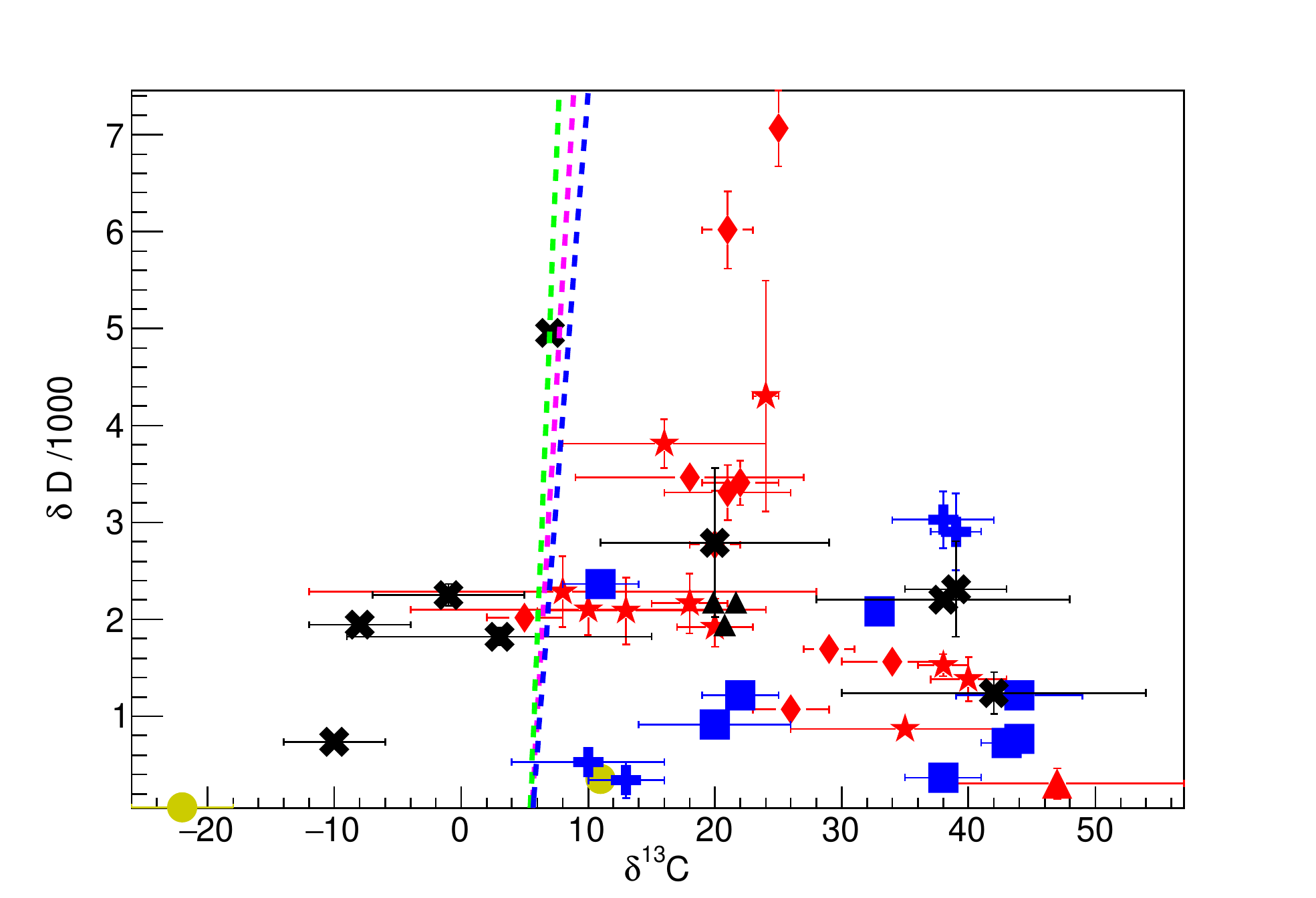}\\
	\includegraphics[width=0.49\textwidth]{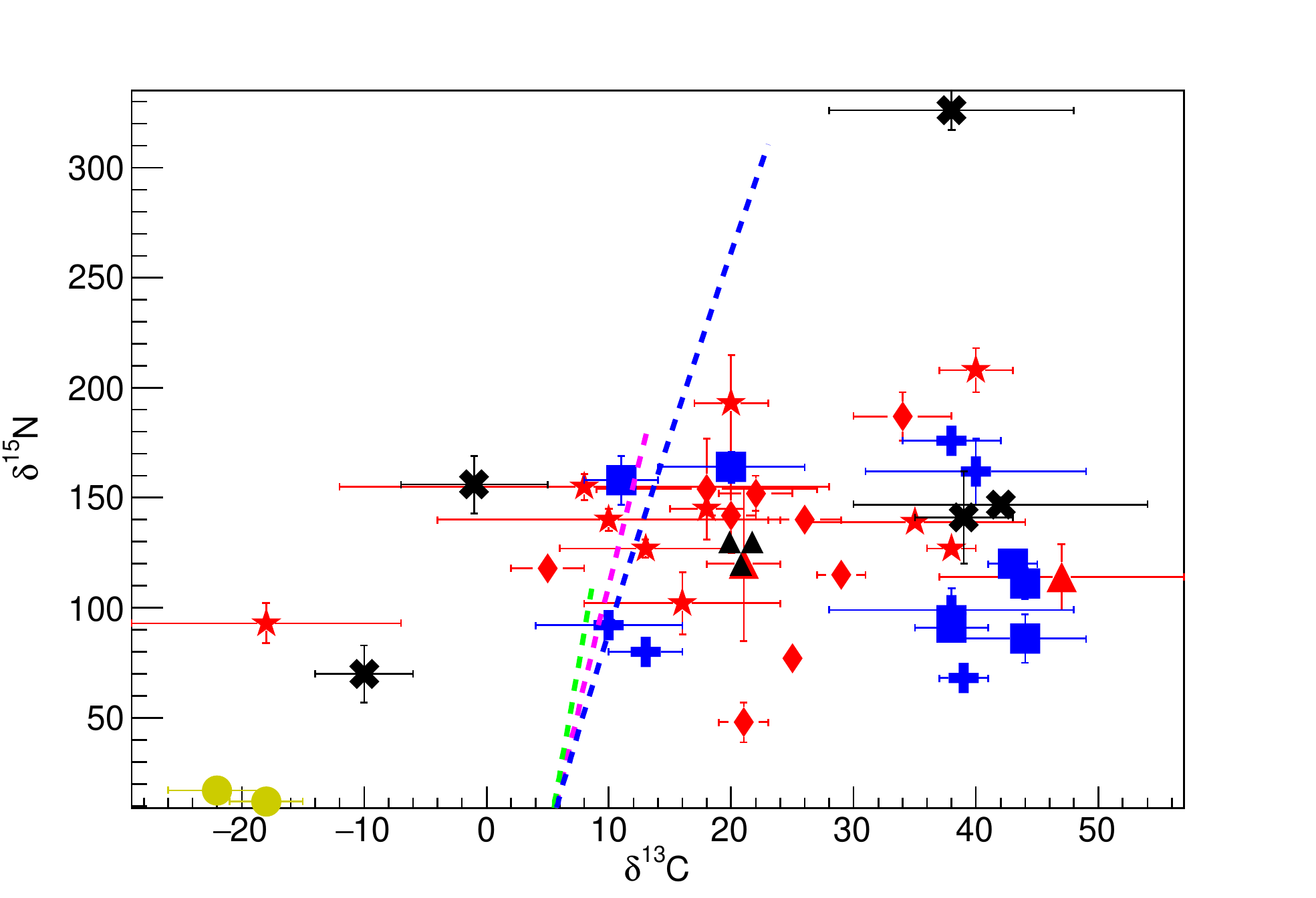}
					\includegraphics[width=0.49\textwidth]{legend-eps-converted-to.pdf}
	\caption{\label{anom_P}
		The abundance ratios compared to the measurements of \citep{elsila12}.  Initial abundances
		are taken to be those of the solar system at formation \citep{lodders09} with isotopic fractions 
		taken from the same reference.  The 
		symbols correspond to the results of Elsila, while the green, purple, and blue lines correspond
		to the results of the network calculation for models D, C, and E respectively.   
		The source meteorite for each measurement is indicated
		by the symbol type, and the meteorite type is indicated
		by the symbol color.  
		The 
		direction of increasing integrated neutrino fluence is also shown as indicated. The black trefoil in each figure represents the arithmetic mean of the data.				
	}
\end{figure}

\begin{figure}
	\centering
	\includegraphics[width=0.49\textwidth]{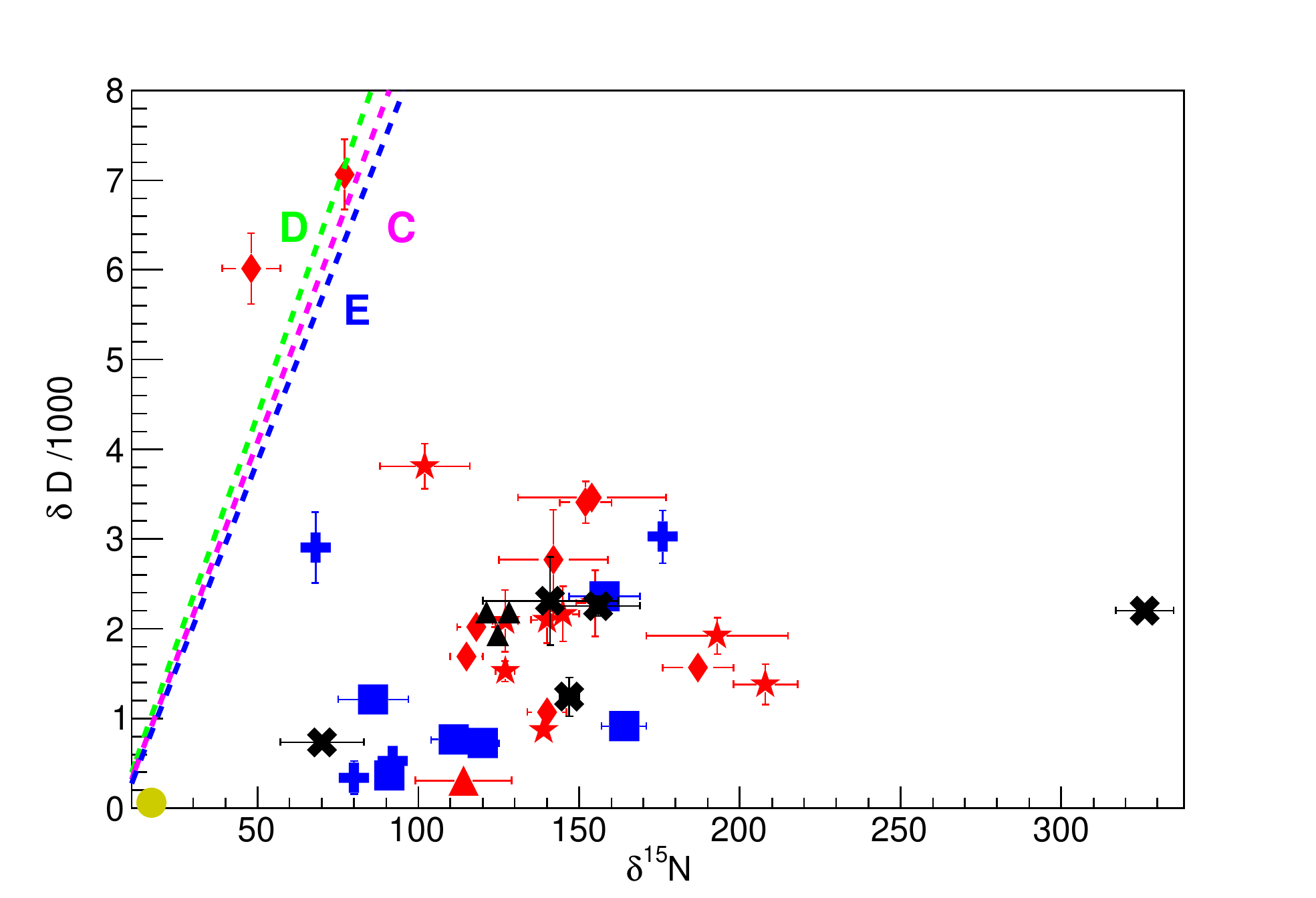}
	\includegraphics[width=0.49\textwidth]{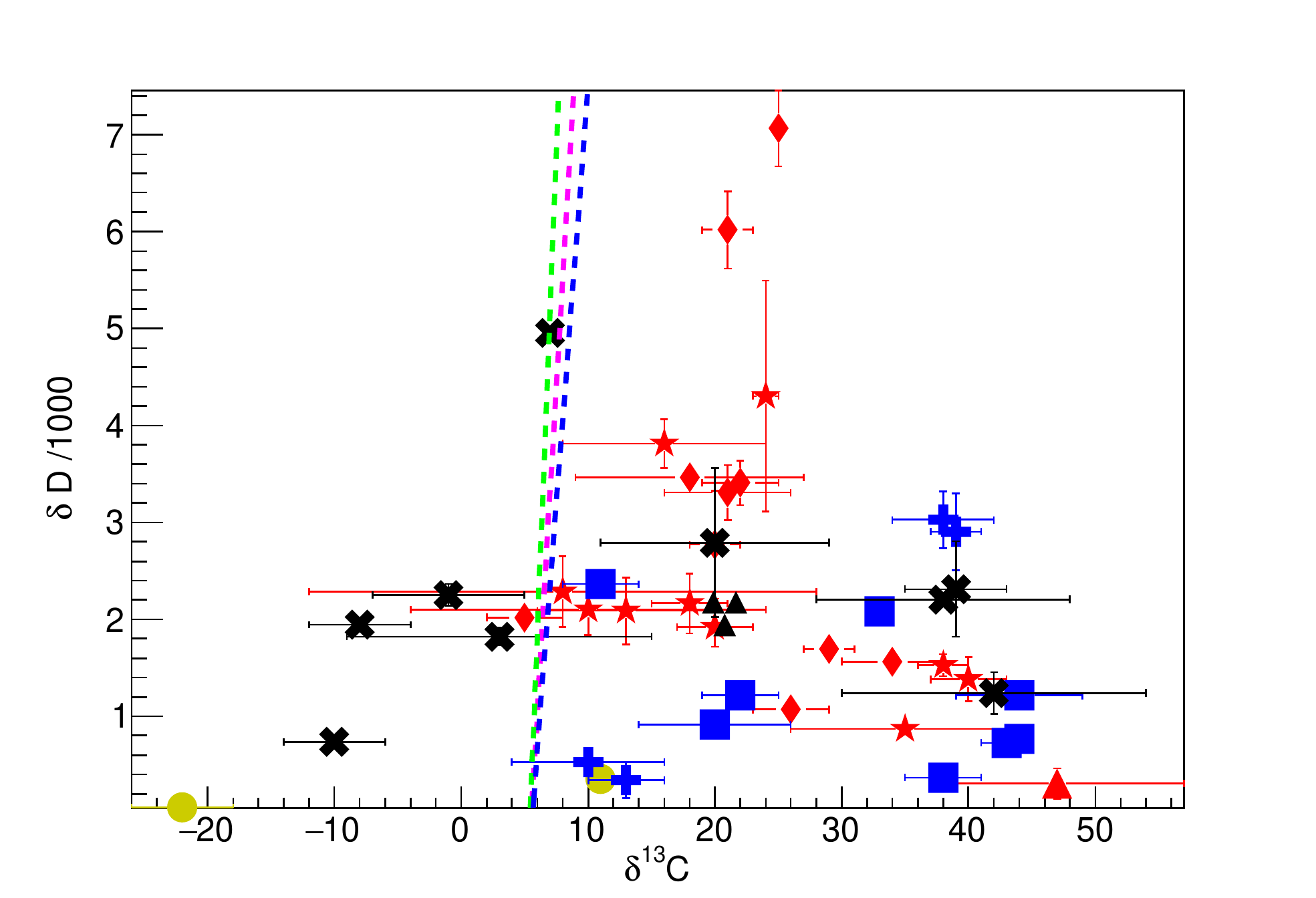}\\
	\includegraphics[width=0.49\textwidth]{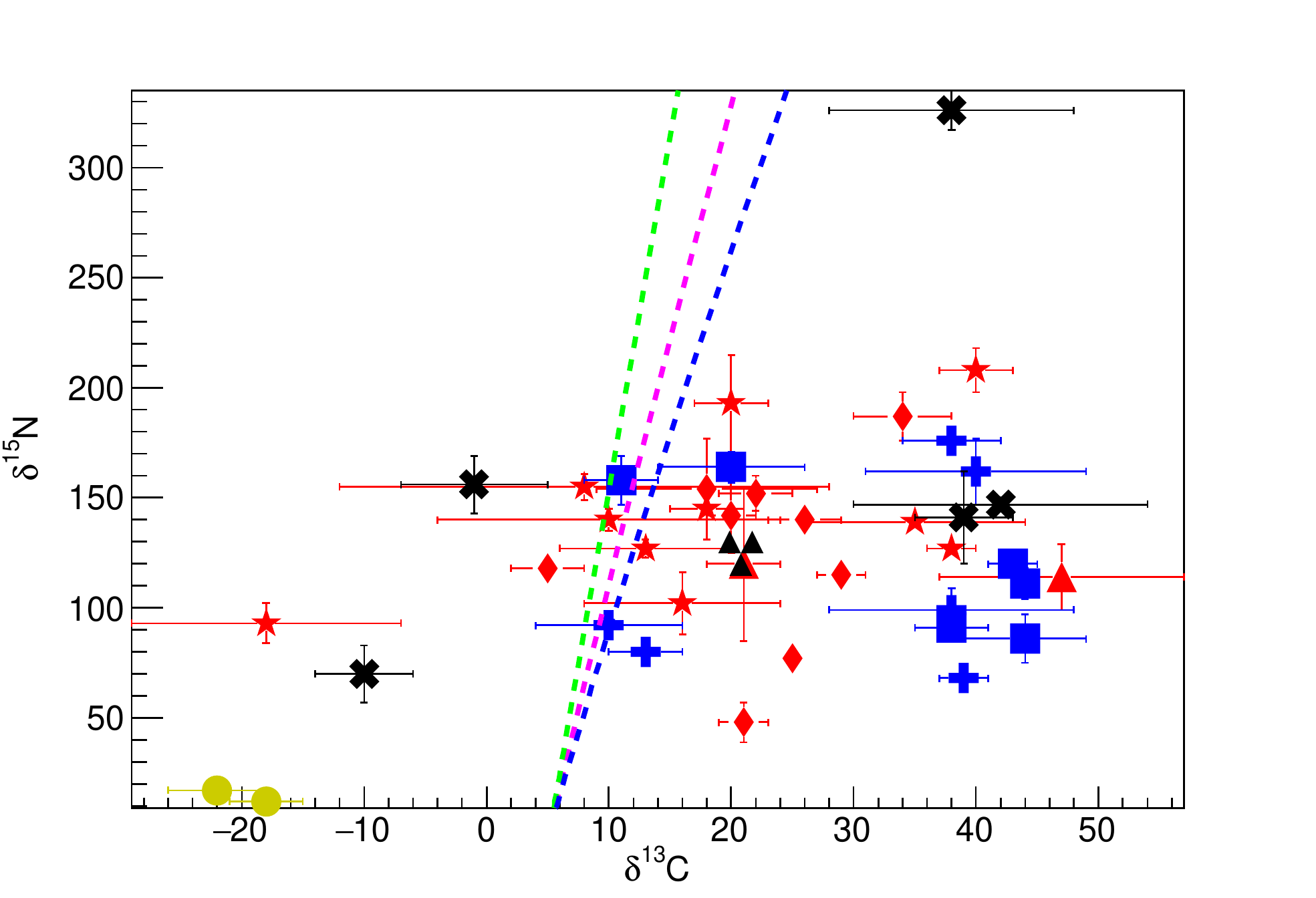}
	\includegraphics[width=0.49\textwidth]{legend-eps-converted-to.pdf}
	\caption{\label{anom_S}
		The abundance ratios compared to the measurements of \citep{elsila12}.  Initial abundances
		are taken to be current suggested solar system values \citep{lodders09} with isotopic fractions 
		taken from the same reference.  The 
		symbols correspond to the results of Elsila, while the green, purple, and blue lines correspond
		to the results of the network calculation for models D, C, and E respectively.   
		The source meteorite for each measurement is indicated
		by the symbol type, and the meteorite type is indicated
		by the symbol color.  
		The 
		direction of increasing integrated neutrino fluence is also shown as indicated. The black trefoil in each figure represents the arithmetic mean of the data.			
	}
\end{figure}

Several things must be noted in comparing the results of the calculations 
with the data. First, the data are amino acid specific, which explains the wide
range of values they exhibit. However, our calculations involve anti-neutrinos 
with energies that far exceed molecular binding energies, so they could not 
possibly be specific to individual amino acids. Thus the results of the 
calculations must be compared to 
some average of the observed values. The arithmetic mean of the data is indicated by the black trefoil mark in Figures \ref{anom_CI} - \ref{anom_S}. 

Second, what our calculations predict is the isotopes produced in reactions 
that would surely destroy the molecules in which the atoms existed before 
the interaction occurred. Thus, if the SNAAP model is correct, what the data 
show is the isotopic ratios in the amino acids that resulted from the 
recombination of the atoms and amino acid fragments that were produced 
in the interactions. 

Would the newly formed amino acids have 
a preferred handedness? The environment in which the detritus from the 
anti-neutrino interactions exists would be ideal for enabling 
autocatalysis~\citep{glavin12,viedma08,blinova14}, which might give the newly formed amino 
acids the same, or even greater, left-handedness than that of the amino acids 
that were not destroyed by the anti-neutrino interactions. Of course, 
recombination could occur on vastly larger time scales than the anti-neutrino
processing.

Third, we note that the anti-neutrino burst could produce a strong $^{15}N$ 
component in the newly formed amino acids. Nitrogen is generally the least 
abundant atom in the H, C, N, and O atoms that are essential for amino acids. The 
anti-neutrinos interacting with the more abundant oxygen, especially if water 
exists in the inclusions, might therefore enhance appreciably the abundance 
of the amino acids over what existed prior to the anti-neutrino processing. 
Furthermore, the amino acids would tend to have enhanced D and $^{15}$N abundances.
Thus the question of the extent to which the newly formed amino acids would 
achieve some chirality becomes very important.

Finally, the recoiling particles from the anti-neutrino induced reactions would affect the abundances of atoms that would be available for recombination. The recoiling nuclei would have energies of only a few eV, so would have little effect on the adjacent molecules. The positrons, however, would have MeVs of energy, so could produce many of the constituents of subsequent amino acid formation, especially if water was abundant.

In the absence of information that would lend guidance to the above considerations, 
the best we could achieve is qualitative agreement between the calculations and the data.

Given these caveats, the computations do compare reasonably to the observed results.  
The calculations are seen to produce the huge variations in the isotopic ratios for D, $^{15}$N, and $^{13}$C. The $\delta$D of several thousand is obtained in our calculations at the same fluence that gives a $\delta^{15}$N of order 150 and a 
$^{13}$C of order 10. Indeed, in Figure 4, with the acknowledgement that the D abundance is likely to be high, the calculations are seen to produce results in fairly good agreement with the indicated data averages at an anti-neutrino fluence within the plausible range.
It appears that the D abundance
is overestimated in this computation, particularly in Figures \ref{anom_P} and \ref{anom_S}.  
However, produced deuterium would be expected to diffuse out of the 
inclusions less readily than would ordinary hydrogen~\citep{ganguly02}, but both would be expected to diffuse out more readily than N or C. This selective diffusion of H might be partially responsible for the very large $\delta$D values in the data. Indeed, this could also explain the fluctuations in the data, as smaller inclusions would be subject to more diffusion than larger ones. 
These two effects have not been included in our calculations.

It is also important to note that amino acids
abundances, isotopic composition, and enantiomeric excesses may have been affected by aqueous alteration processes \citep{sephton04,glavin09}.  This could be responsible for the 
scatter in the measured isotopic ratios indicated in the figures.  Thus, the scatter 
in measured values - and possibly the values themselves - may be loosely correlated with the type of 
meteorite from which the amino acids were extracted.  For example,
measurements of $\delta$D appear to cluster near lower values for CR3 and, to a lesser extent, the CR2 meteorites.  Further,
the CR3 meteorites exhibit a lower range of values of $\delta^{13}$C than the CM types.  These differences
in scatter and ranges of values seem to be consistent with
the fact that the CR type meteorites are believed to 
contain the most primitive organic material (\citep{glavin09}).

The model presented here does not predict or account for 
any aqueous alteration processes.  However, we may be able to 
draw a loose comparison to the more primitive meteorites 
studied.  In Figure \ref{anom_CI}, for example, we see that 
our model E is able to produce lower values of $\delta$D, which 
may compare more favorably to the more primitive meteorites.  Likewise, all models are able to predict lower values of $\delta^{13}$C consistent with the more primitive meteorites. 
\subsection{Composition of Meteoric Inclusions}
Calculations were  also performed in which inhomogeneities in the inclusions were assumed.  
The evaluation of the
effects of water on any nuclear processes in the meteorite 
is rudimentary as the parent body history may vary from one 
body to the next.  The current model
would have to reflect the history of a large variety of meteroids and comets, with potentially large variations in 
the amount of water contained within them~\citep{alexander18,marty16,jewitt16}. Thus we assume only
a single case with inclusions containing water
with an assumed water mass fraction of 10\%. This might correspond to 
a body rich in water, but not necessarily the most water-rich
body but not necessarily the largest enrichment thought possible
~\citep{marty16}. 
In this evaluation, the water could be either liquid or ice,
since 
shifts in nuclear species are independent of the chemical 
phase.  Water could take the form of small pockets or ``bubbles'' 
in the meteoric inclusions. 
As stated previously, the nuclear processes studied are independent of
the actual molecular species being studied and depend only on the isotopic (nuclear) composition.  Any chemical processing would be independent and in addition to the processes studied here.

The isotopic fractions assumed for the inclusions are 
shown in Table \ref{water_inc}.    The individual isotopic fractions were assumed to be that of the Orgueil meteorite
as listed in \citep{lodders09}.  Initial isotopic ratios (e.g. D/H)
were assumed to be those of Table 10 of the same reference.  
(Note that mass fractions,
and not isotopic abundance fractions are indicated in Table \ref{water_inc}.)  Although this is a high fraction of water,
it was used to obtain a qualitative estimate of the effects of
aqueous environments on the meteoric abundances.

In order to evaluate isotopic abundances in water inclusions, we assumed that the anti-neutrino flux will initially alter the abundances within
the inclusion.  However, because of the large mean-free-path 
of the neutrons, we assumed a subsequent neutron flux of a
homogeneously mixed carbonaceous chrondrite with a composition consistent with that of the Orgueil meteorite. Thus the neutron abundance and energy for water
inclusions is assumed to be the same as that of the meteoroid.  Likewise the fraction
of neutrons captured in any specific reaction is the same as average neutron
capture rates based on the average composition of meteoric material.  Locally (i.e.,
within the inclusion), the fractional amount of an isotopic species undergoing a reaction is the same as the meteoric average.  
The absolute amount, however, depends
on the local abundance.  

\begin{figure}
	\centering
	\includegraphics[width=0.49\textwidth]{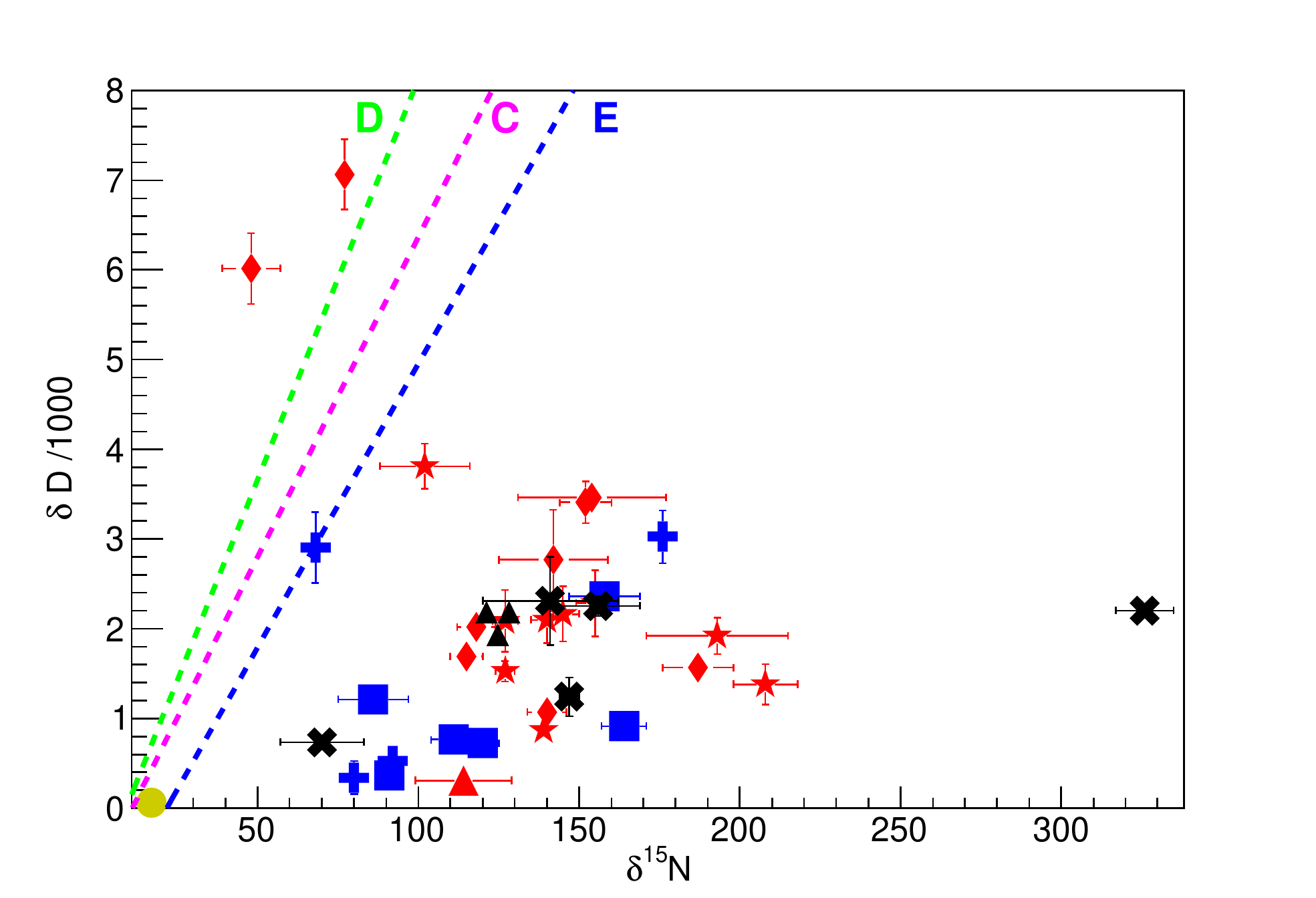}
	\includegraphics[width=0.49\textwidth]{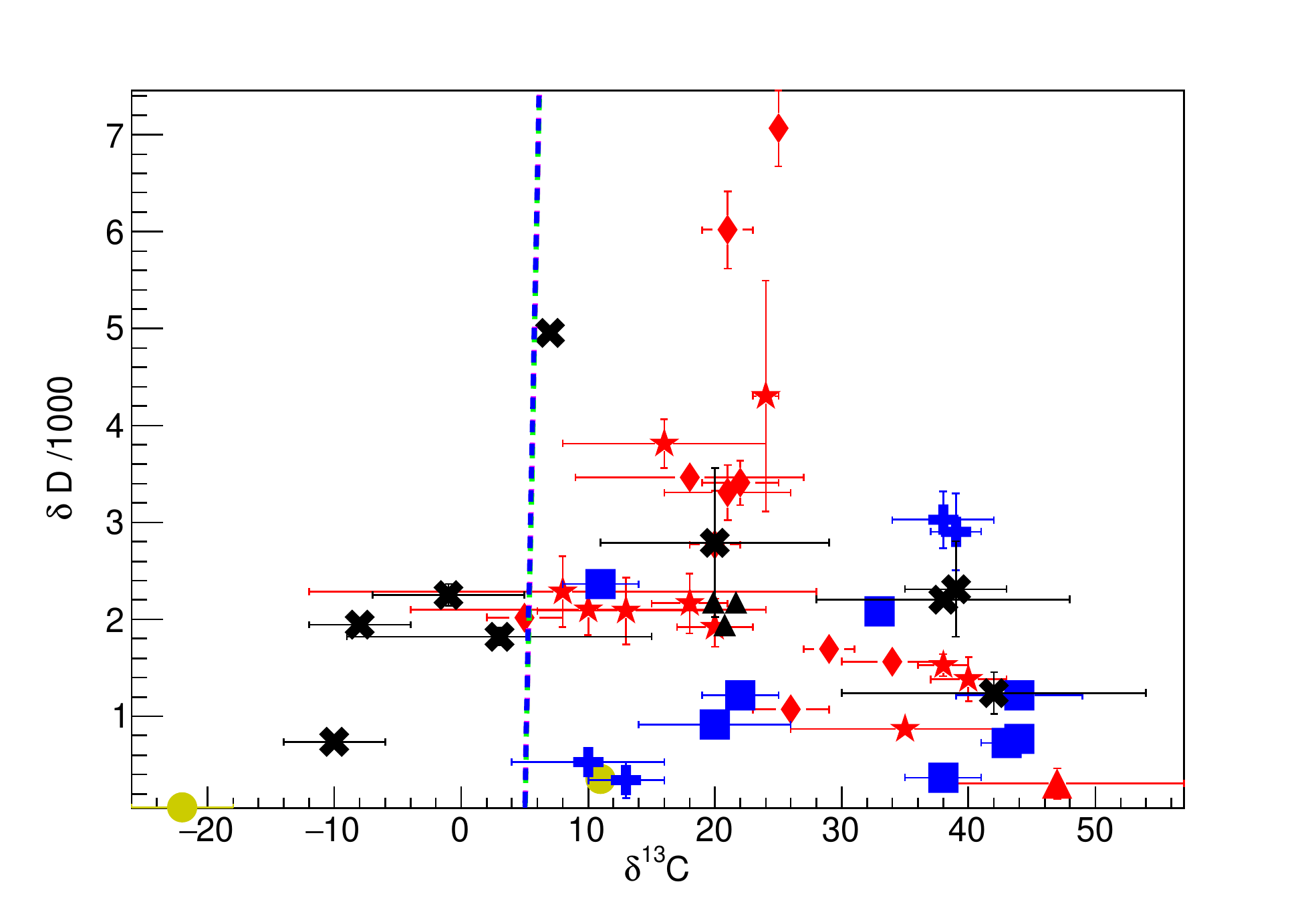}\\
	\includegraphics[width=0.49\textwidth]{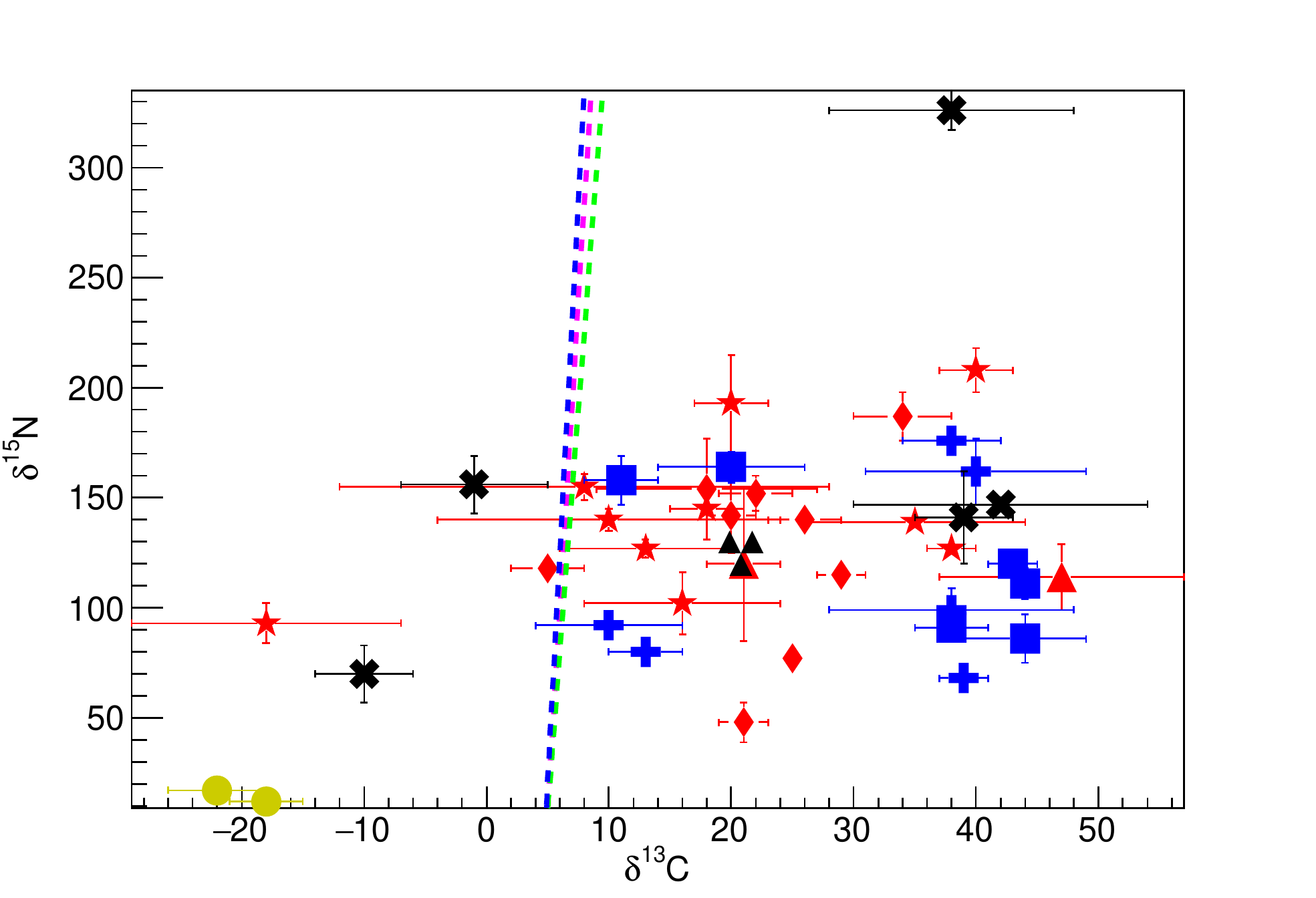}
	\includegraphics[width=0.49\textwidth]{legend-eps-converted-to.pdf}
	\caption{\label{anom_W}
		The abundance ratios compared to the measurements of \citep{elsila12}.  Initial abundances
		are taken to be those of an inclusion with abundances given in the ``water'' column Table \ref{water_inc}.  The 
		symbols correspond to the results of Elsila, while the green, purple, and blue lines correspond
		to the results of the network calculation for models D, C, and E respectively.   
		The source meteorite for each measurement is indicated
		by the symbol type, and the meteorite type is indicated
		by the symbol color.  
		The 
		direction of increasing integrated neutrino fluence is also shown as indicated. The black trefoil in each figure represents the arithmetic mean of the data.				
	}
\end{figure}

The results for this analysis are shown in Figure \ref{anom_W}.  It can be seen that the results are very similar to those of
the Orgueil meteorite, though the values of $\delta$D appear somewhat higher.
None the less the same qualitative agreement with the Orgueil results as seen without the water inclusion is observed.  

\begin{figure}
	\centering
	\includegraphics[width=0.49\textwidth]{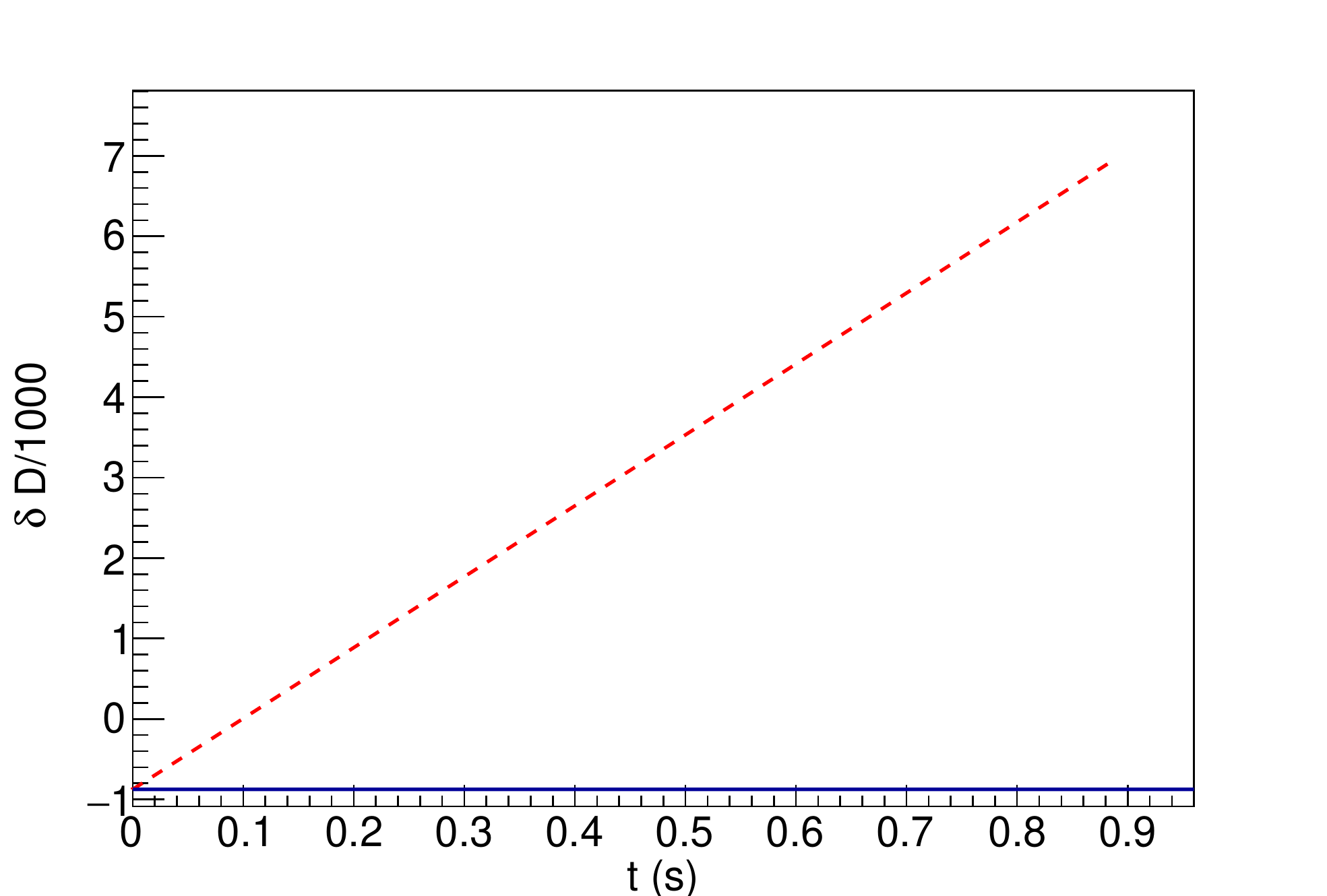}
	\includegraphics[width=0.49\textwidth]{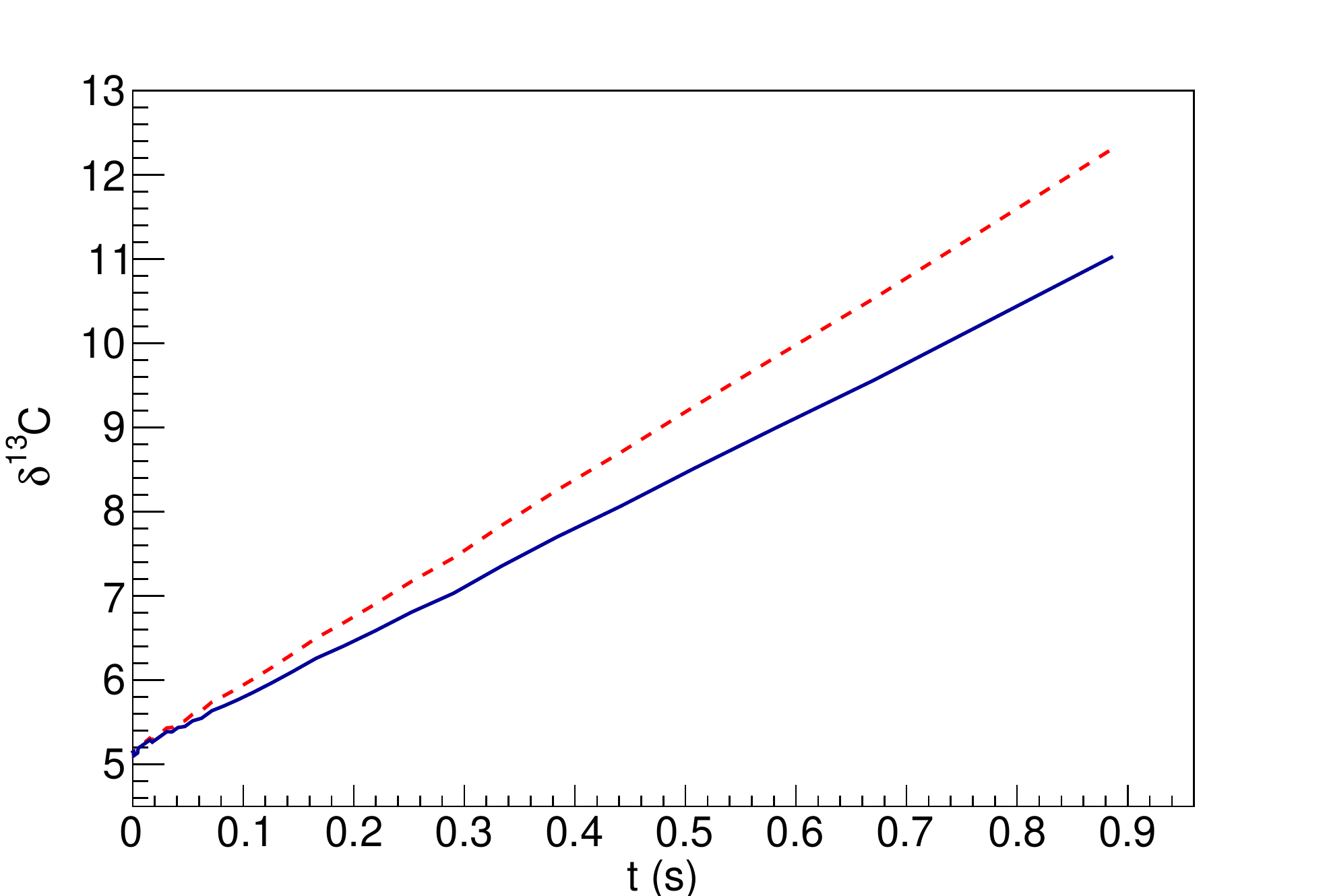}\\
	\includegraphics[width=0.49\textwidth]{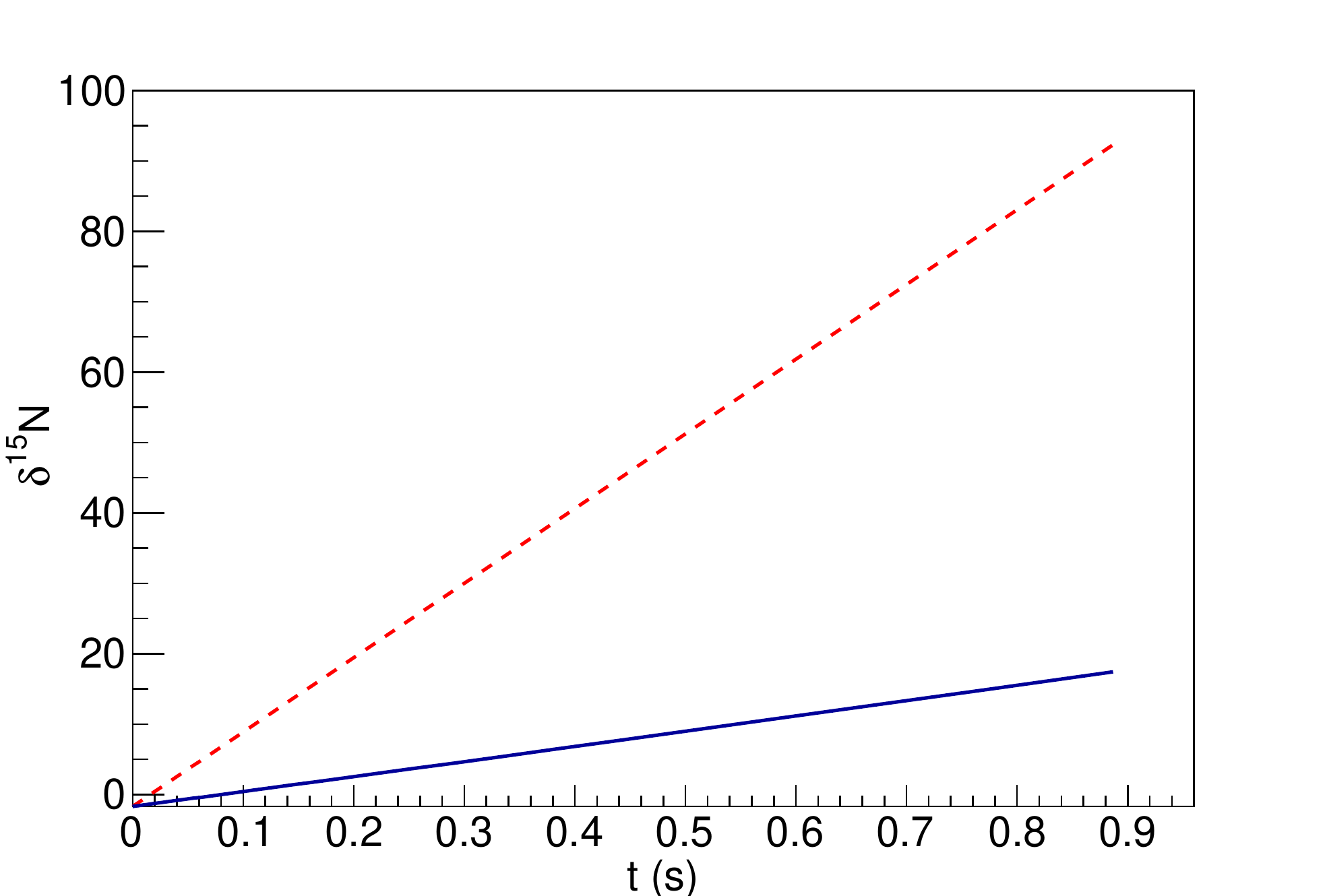}
	\caption{\label{delta_compare}
		The isotopic ratios $\delta D$, $\delta^{13}C$, and $\delta^{15}N$ as a function of time (which corresponds to integrated neutrino flux) for cases in with neutron captures (dashed line) and in which subsequent neutron captures are disabled.	In this
		figure, the neutrino flux corresponding to model E is assumed for an initial composition of the
		solar system at the time of formation.  		
	}
\end{figure}

Some understanding can be gained from this result by examining the significance of neutron captures, which is shown in Figure 
\ref{delta_compare}.  Here, the values of $\delta D$, $\delta^{13}C$, and $\delta^{15}N$ are shown as a function of time in the
neutrino flux if neutron captures are enabled or disabled.  In each situation, an isotopic shift is induced either via the neutral current reactions:
\begin{eqnarray}
\label{o_nu}
^{16}O + \nu(\bar{\nu})&\rightarrow ^{15}N + ^{1}H + \nu(\bar{\nu})\\
\label{n_nu}
^{14}N + \nu(\bar{\nu})&\rightarrow ^{13}C + ^{1}H + \nu(\bar{\nu})\\
\label{he_nu}
^{3}He + \nu(\bar{\nu})&\rightarrow ^{2}H + ^{1}H + \nu(\bar{\nu})
\end{eqnarray}
or via neutron capture:
\begin{eqnarray}
\label{n_n}
^{14}N + n &\rightarrow ^{15}N\\
\label{c_n}
^{12}C + n &\rightarrow ^{13}C\\
\label{h_n}
H + n &\rightarrow ^2H
\end{eqnarray}
 In Figure \ref{delta_compare}, we see that disabling neutron captures results in effectively disabling all deuterium production, while only slightly
 suppressing $^{13}$C and suppressing $^{15}$N production somewhat. Thus deuterium is produced almost entirely via the reaction given by Equation
 \ref{h_n}.  The production of $^{15}$N proceeds primarily through
 the reaction given by Equation \ref{n_n} with a contribution from
 the reaction given by Equation \ref{o_nu}.  The production of $^{13}$C 
 proceeds primarily via the reaction given by Equation \ref{n_nu}.
 
The inclusion of water increases
 the initial H  and D mass fractions as seen in Table \ref{water_inc}, while the remaining mass fractions
 change by smaller absolute amounts. 
 
It is assumed that  the overall average neutron flux does not change 
  even
 if water inclusions are assumed as most of the neutrons are produced in the surrounding 
 rock.  This means that the fraction of H converted to deuterium and the fraction of
 $^{14}$N converted to $^{15}$N does not change.  However, since the absolute amount
 of H is larger, the total amount of deuterium produced is also larger.  
 
 In the case of $^{15}$N, there is some dependence on neutron captures, but also a small dependence on neutral current interactions in producing $^{15}$N via the reaction 
 \ref{o_nu}.
 There is little change in the overall ratio, however, because the meteoric material is very rich in oxygen regardless of the inclusion composition and the overall
 oxygen abundance changes little.  
 Likewise, the N/O ratio in this sample is identical to that of the Orgueil 
 meteorite assumed in Figure 
 \ref{anom_CI}.  
 
 It's also worth noting the timescales in Figure \ref{delta_compare}.  While Figures \ref{anom_CI} -- \ref{anom_W} show
 the correlations between isotopic ratios, the time dependence (e.g., which point in the figures corresponds to how much time has elapsed in the model)
 is not
 shown explicitly in those figures, while it
 is present in Figure \ref{delta_compare}.  It is noted that the ratios 
 calculated coincide with the measurements in less than 1 s, which is
 the approximate duration of the neutrino flux in a NS merger \citep{rosswog03}, one of the postulated sites of this model \citep{snaap3,snaap4,snaap5,snaap6}.

Weak interactions would likely be responsible 
for a wide variety of differences in isotopic abundance ratios in the same meteorites,
specifically $D/H$, $^{13}$C/$^{12}$C, $^{15}$N/$^{14}$N, $^{7}$Li/$^{6}$Li and $^{10}$B/$^{11}$B.  While Li and
B are not present in meteoric amino acids, neutrino interactions in the surrounding rock might be thought to induce similar isotopic abundance changes.

Lithium and boron isotopic ratios have also been studied in meteorites
	in which amino acid enantiomeric excesses and isotopic enhancements have
	been found~\citep{sephton13,gyngard07,marhas02,mcdonough03}.  
	In these studies, variations in the  
	$^7$Li and $^{10}$B isotopic ratios have been observed.  
	Of course, any chemical, thermal, and physical
	processes for Li and B are independent of any amino acid processing and 
	chemistry.  Thus, Li and B processing should be treated independently.  However, any meteoroid which contains organic
	material will also contain other elements, including Li and B.  Because of
	this, any processing of amino acids at the nuclear level also occurs
	for other elements in the same body.

While we have considered processing of Li
and B isotopic ratios, we have found the uncertainties in our
calculations to be too great to render an effective, conclusive
calculation.  We hope to explore processing of Li and B
in subsequent studies.

While the evaluation of abundances using the initial meteoritic abundances of the Orgueil meteorite \citep{lodders09} most closely 
matches current measurements, those initial abundances
used were those taken after any presumed processing has occurred.  
It should also be noted that
enrichment in L-enantiomers from meteorites are 
somewhat disputed as terrestrial contamination may have 
influenced the results.
A more accurate evaluation
could be done with abundances of a primitive meteorite, 
or better yet, from samples returned to Earth by Hayabusa2 \citep{wa19, ki19, su19, ja19}.

\section{Conclusions}
In previous works \citep{snaap1,snaap2,snaap3,snaap4,snaap5,snaap6, boyd18}, the SNAAP model 
was shown to offer a possible explanation of amino acid chiral selection in 
stellar environments via parity-violating weak interactions.  The quantitative 
results have suggested that this model could explain
the selection of left-handed amino acids in stellar meteorites.

This work shows that this model is not only capable of inducing amino acid
chiral selection in meteorites but also produces shifts in isotopic abundance ratios which are comparable
to those observed in the same meteorites.  
Furthermore, the required electron anti-neutrino fluences required in the two situations are comparable.

We studied three scenarios of the presumed
neutrino flux and energy spectrum in a neutron-star merger
model.  Using a nuclear reaction network, we 
examined the possible change in isotopic abundance ratios
in meteoric material.  The model used here simultaneously
predicts isotopic shifts in both organic material and
the surrounding rock.  We emphasize that the predictions
in the figures show time evolution of predicted abundance
ratios.  (That is, each dot in the shown results 
corresponds to a different integrated flux.) Thus,
an observed result should correspond to only one of
any of the dots in the figures.  While we do not attempt to
replicate the scatter in the measurements or any
subsequent chemistry, we do note that many of our
predictions appear to fall within the locus of
observed results.  

Of the models studied, Model C of
the~\cite{rosswog03} merger model seems to produce
abundance distributions most consistent with observations.
Of the initial abundance distributions studied,
initial abundances matching those of the early solar 
system result in calculations similar to observed results, though the scatter in measurements is significant.

Though observed values have a large amount of uncertainty, even within a single
parent body, we note that our computations here are able to provide results that
are within the range of observations.  Additionally, because the
anti-neutrino fluence varies from one meteoroid to the next, the range of possible
values of isotopic ratios between meteoritic measurement can be explained.  While
we have not explored the chemistry involved, we note that this may also contribute
to isotopic variations between models.

The study of weak interactions in selecting amino acid chirality is
is still quite young, and there are many directions to take.  In the case of 
meteoric abundances, future work will concentrate on careful evaluation of individual
astrophysical sites in which this model may work as well as further signatures of this model.

One might also consider the effects on
the production of $^{27}$Al in similar
studies as many of the isotopic ratio measurements pertained to studies of SiC grains. Reactions on Si in such grains might also be
responsible for $^{27}$Al production. Future studies of meteoric samples might want to include this possibility.

It will be especially interesting if Hayabusa2 returns to Earth with sufficiently 
large samples from asteroid 162173 Ryugu to determine the enantiomeric excesses of the amino acids and the
variations in isotopic 
ratios of the five elements discussed in this paper.
\section*{Acknowledgments}
MAF is supported by a Moore Foundation grant \#7799, the Fulbright Foundation, and by
 the NAOJ visiting professor program.  
 TK is supported in part by Grants-in-Aid for Scientific Research of JSPS (15H03665, 17K05459).
 TS is supported in part by JSPS KAKENHI Grant 
 Number JP19K03855. MAF and SC both acknowledge support from a visiting professorship at 
 Beihang University.The authors would also like to thank the anonymous referee for several 
 excellent comments, which have greatly improved this manuscript.

\newcommand{\prl}{Phys. Rev. Lett.}
\newcommand{\prc}{Phys. Rev. C}
\newcommand{\apj}{Astroph. J.}
\newcommand{\ssr}{Space Science Rev.}

\bibliographystyle{astrobiology}
\bibliography{references}
\section*{Tables}
\begin{table}[h!]
	\centering
	\caption{\label{models} Neutrino temperatures (in MeV) and 
		luminosities (10$^{53}$ erg/s) for the three models used to
		compute the isotopic anomolies of the amino acids in this work.  
		These parameters are taken from
		\citep{rosswog03}.}
	\begin{tabular}{|l|c|c|c|}
		\hline
		~&\multicolumn{3}{c|}{\textbf{Models}}\\\hline
		~&C&D&E\\\hline
		T$_{\nu_e}$& 4.2 & 4.1 & 5.0\\\hline
		T$_{\bar{\nu}_e}$& 5.5 & 5.0 & 6.6\\\hline
		T$_{\nu_x}$ & 8.9 & 6.6 & 9.1\\\hline
		L$_{\nu_e}$ & 0.5 & 0.3& 1.5\\\hline
		L$_{\bar{\nu}_e}$ & 1.5 & 0.9 & 2.25 \\\hline
		L$_{\nu_x}$ & 0.25 & 1 & 0.75\\\hline
	\end{tabular}
\end{table}
\newpage
\begin{table}[t!]
	\centering
	\caption{\label{water_inc} Assumed initial mass fractions
	of isotopes for the carbonaceous chondrite \citep{lodders09} and water inclusions in
	meteorites.}
	\begin{tabular}{|c|c|c||c|c|c|}
		\hline
	    \textbf{Isotope} &\textbf{CC} & \textbf{Water} 
	    & \textbf{Isotope} & \textbf{CC} & \textbf{Water}\\
	    \hline
	    \hline
		\textbf{$^1$H} & 3.74$\times$10$^{-2}$ & 3.77$\times$10$^{-2}$ & \textbf{$^{12}$C} & 6.58$\times$10$^{-2}$ & 5.96$\times$10$^{-2}$\\
		\textbf{$^2$H} & 1.45$\times$10$^{-6}$ & 1.45$\times$10$^{-6}$ & \textbf{$^{13}$C} & 8.00$\times$10$^{-4}$ & 7.25$\times$10$^{-4}$ \\
		\textbf{$^3$He} & 2.18$\times$10$^{-12}$  & - & \textbf{$^{14}$N} & 5.62$\times$10$^{-3}$ & 5.10$\times$10$^{-3}$\\
		\textbf{$^4$He} &  1.75$\times$10$^{-8}$ & - & \textbf{$^{15}$N} & 2.21$\times$10$^{-5}$ & 2.01$\times$10$^{-5}$\\
		\textbf{$^6$Li} &  1.84$\times$10$^{-7}$ & - & \textbf{$^{16}$O} & 0.888 & 0.894\\
		\textbf{$^7$Li}& 2.66$\times$10$^{-6}$ & - & \textbf{$^{17}$O}  & 3.50$\times$10$^{-4}$ & 3.52$\times$10$^{-4}$\\
		\textbf{$^9$Be} & 4.02$\times$10$^{-8}$ & - & \textbf{$^{18}$O} & 2.00$\times$10$^{-3}$ & 2.02$\times$10$^{-3}$\\
		\textbf{$^{10}$B} & 2.72$\times$10$^{-7}$ & - & \textbf{$^{19}$F}  & 1.11$\times$10$^{-4}$ & -\\
		\textbf{$^{11}$B} & 1.21$\times$10$^{-6}$ & - & \textbf{$^{20}$Ne} & 3.18$\times$10$^{-10}$ & -\\\cline{1-3}
		\multicolumn{3}{c||}{~} & \textbf{$^{21}$Ne}  & 8.01$\times$10$^{-13}$ & -\\
		\multicolumn{3}{c||}{~} & \textbf{$^{22}$Ne}  & 2.58$\times$10$^{-11}$ & -\\
		\cline{4-6}
	\end{tabular}
\end{table}
\clearpage
\section*{Figure Legends}
Figure \ref{w_net}: The weak interaction network used to evaluate isotopic abundance distributions in amino acid formation.\\
\\
Figure \ref{anom_CI}:
		The abundance ratios compared to the measurements of \citep{elsila12}.  Initial abundances
		are taken to be those of the Orgueil meteorite \citep{lodders09} with isotopic fractions 
		taken from the same reference.  The 
		symbols correspond to the results of Elsila, while the green, purple, and blue lines correspond
		to the results of the network calculation for models D, C, and E respectively.   
		The source meteorite for each measurement is indicated
		by the symbol type, and the meteorite type is indicated
		by the symbol color.  
		The 
		direction of increasing integrated neutrino fluence is also shown as indicated. The black trefoil in each figure represents the arithmetic mean of the data.	\\
\\		
Figure \ref{anom_P}:
		The abundance ratios compared to the measurements of \citep{elsila12}.  Initial abundances
		are taken to be those of the solar system at formation \citep{lodders09} with isotopic fractions 
		taken from the same reference.  The 
		symbols correspond to the results of Elsila, while the green, purple, and blue lines correspond
		to the results of the network calculation for models D, C, and E respectively.   
		The source meteorite for each measurement is indicated
		by the symbol type, and the meteorite type is indicated
		by the symbol color.  
		The 
		direction of increasing integrated neutrino fluence is also shown as indicated. The black trefoil in each figure represents the arithmetic mean of the data.		\\
\\		
Figure \ref{anom_S}:
		The abundance ratios compared to the measurements of \citep{elsila12}.  Initial abundances
		are taken to be current suggested solar system values \citep{lodders09} with isotopic fractions 
		taken from the same reference.  The 
		symbols correspond to the results of Elsila, while the green, purple, and blue lines correspond
		to the results of the network calculation for models D, C, and E respectively.   
		The source meteorite for each measurement is indicated
		by the symbol type, and the meteorite type is indicated
		by the symbol color.  
		The 
		direction of increasing integrated neutrino fluence is also shown as indicated. The black trefoil in each figure represents the arithmetic mean of the data.	 \\
\\		
Figure \ref{anom_W}:
		The abundance ratios compared to the measurements of \citep{elsila12}.  Initial abundances
		are taken to be those of an inclusion with abundances given in the ``water'' column Table \ref{water_inc}.  The 
		symbols correspond to the results of Elsila, while the green, purple, and blue lines correspond
		to the results of the network calculation for models D, C, and E respectively.   
		The source meteorite for each measurement is indicated
		by the symbol type, and the meteorite type is indicated
		by the symbol color.  
		The 
		direction of increasing integrated neutrino fluence is also shown as indicated. The black trefoil in each figure represents the arithmetic mean of the data.	 \\
\\		
Figure \ref{delta_compare}:
		The isotopic ratios $\delta D$, $\delta^{13}C$, and $\delta^{15}N$ as a function of time (which corresponds to integrated neutrino flux) for cases in with neutron captures (dashed line) and in which subsequent neutron captures are disabled.	In this
		figure, the neutrino flux corresponding to model E is assumed for an initial composition of the
		solar system at the time of formation.  \\
\\		
	\clearpage
\section*{Figures}
\clearpage

\clearpage
\clearpage
\clearpage

\clearpage
\end{document}